\documentclass[journal,draftclsnofoot,onecolumn,12pt,a4paper]{IEEEtran}
\IEEEoverridecommandlockouts
\usepackage{amssymb}
\usepackage{amsfonts}
\usepackage{epstopdf}
\usepackage{graphicx}
\usepackage{stmaryrd}
\usepackage{amssymb,amsmath}
\usepackage{multirow,url}
\usepackage{color,cite}
\usepackage[bookmarks=false]{hyperref}
\usepackage{comment}
\usepackage{subcaption}
\usepackage{diagbox} 

\ifodd 1
\newcommand{\com}[1]{\textbf{\color{red} (COMMENT: #1)}} 
\newcommand{\comg}[1]{\textbf{\color{green} (COMMENT: #1)}}
\newcommand{\response}[1]{\textbf{\color{magenta} (RESPONSE: #1)}} 
\else

\newcommand{\com}[1]{}
\newcommand{\comg}[1]{}
\newcommand{\response}[1]{}
\fi
\ifodd 0
\newcommand{\referred}[1]{\textcolor{red}{RefPaper: #1}} 
\else
\newcommand{\referred}[1]{}
\fi

\ifodd 0
\newcommand{\changeblue}[1]{\textcolor{blue}{Modified: #1}} 
\else
\newcommand{\changeblue}[1]{}
\fi

\begin{document}

\title{Practical Power-Balanced Non-Orthogonal Multiple Access}


\author{Haoyuan~Pan,~\IEEEmembership{Student Member,~IEEE,}£¬~Lu~Lu,~\IEEEmembership{Member,~IEEE,}
       and~Soung~Chang~Liew,~\IEEEmembership{Fellow,~IEEE}
\thanks{H. Pan and S. C. Liew are with the Department of Information Engineering, The Chinese University of Hong Kong, Hong Kong.
Email: \{ph014, soung\}@ie.cuhk.edu.hk. L. Lu is with the Institute of Network Coding, The Chinese University of Hong Kong, Hong Kong.
Email: lulu@ie.cuhk.edu.hk.
}
}

\maketitle

\begin{abstract}
This paper investigates practical 5G strategies for power-balanced non-orthogonal multiple access (NOMA). By allowing multiple users to share the same time and frequency, NOMA can scale up the number of served users and increase spectral efficiency compared with existing orthogonal multiple access (OMA). Conventional NOMA schemes with successive interference cancellation (SIC) do not work well when users with comparable received powers transmit together. To allow power-balanced NOMA (more exactly, near power-balanced NOMA), this paper investigates a new NOMA architecture, named Network-Coded Multiple Access (NCMA). A distinguishing feature of NCMA is the joint use of physical-layer network coding (PNC) and multiuser decoding (MUD) to boost NOMA throughputs. We first show that a simple NCMA architecture in which all users use the same modulation, referred to as \emph{rate-homogeneous NCMA}, can achieve substantial throughput improvement over SIC-based NOMA under near power-balanced scenarios. Then, we put forth a new NCMA architecture, referred to as \emph{rate-diverse NCMA}, in which different users may adopt different modulations commensurate with their relative SNRs. A challenge for rate-diverse NCMA is the design of a channel-coded PNC system. This paper is the first attempt to design channel-coded rate-diverse PNC.  Experimental results on our software-defined radio prototype show that the throughput of rate-diverse NCMA can outperform the state-of-the-art rate-homogeneous NCMA by 80\%. Overall, rate-diverse NCMA is a practical solution for near power-balanced NOMA.
\end{abstract}

\begin{IEEEkeywords}
Network-coded multiple access, physical-layer network coding, multi-user detection, non-orthogonal multiple access, implementation
\end{IEEEkeywords}


\IEEEpeerreviewmaketitle



\newpage

\section{Introduction}\label{sec:intro}
Non-orthogonal multiple access (NOMA) is a promising technique to increase spectral efficiency in 5G cellular networks. For uplink NOMA, multiple users transmit simultaneously to a base station (BS) with non-orthogonal signaling (specifically, all users transmit at the same time, in the same frequency band, and without using different code signatures)\referred{NOMAfor5G,NOMAVTC13,powerdomainNOMA}\cite{NOMAfor5G,NOMAVTC13,powerdomainNOMA}. By allowing multiple users to share the same time and frequency, NOMA can scale up the number of served users and increase spectral efficiency compared with existing orthogonal multiple access (OMA), e.g., TDMA, FDMA, and OFDMA \referred{Verdubook}\cite{Verdubook}.

Successive interference cancellation (SIC) has been studied widely as a NOMA technique \referred{NOMAfor5G,Verdubook}\cite{NOMAfor5G,Verdubook}. In SIC-based NOMA, different end users are clustered into small groups. The users within a group transmit at the same time using the same frequency and waveform (i.e., using NOMA). The users are grouped in such a way that within each group, the users' powers received at the BS are widely different. The large received power differences among the users are key to the inner workings of SIC. It has further been suggested that NOMA should pair a strong user with a weak user in a two-user group (namely, an SIC group) to improve the overall multiuser decoding (MUD) system throughput \referred{NOMAfor5G,NOMAdynamicgroup}\cite{NOMAfor5G,NOMAdynamicgroup}.

\begin{figure}[t]
\centering
\includegraphics[width=0.45\textwidth]{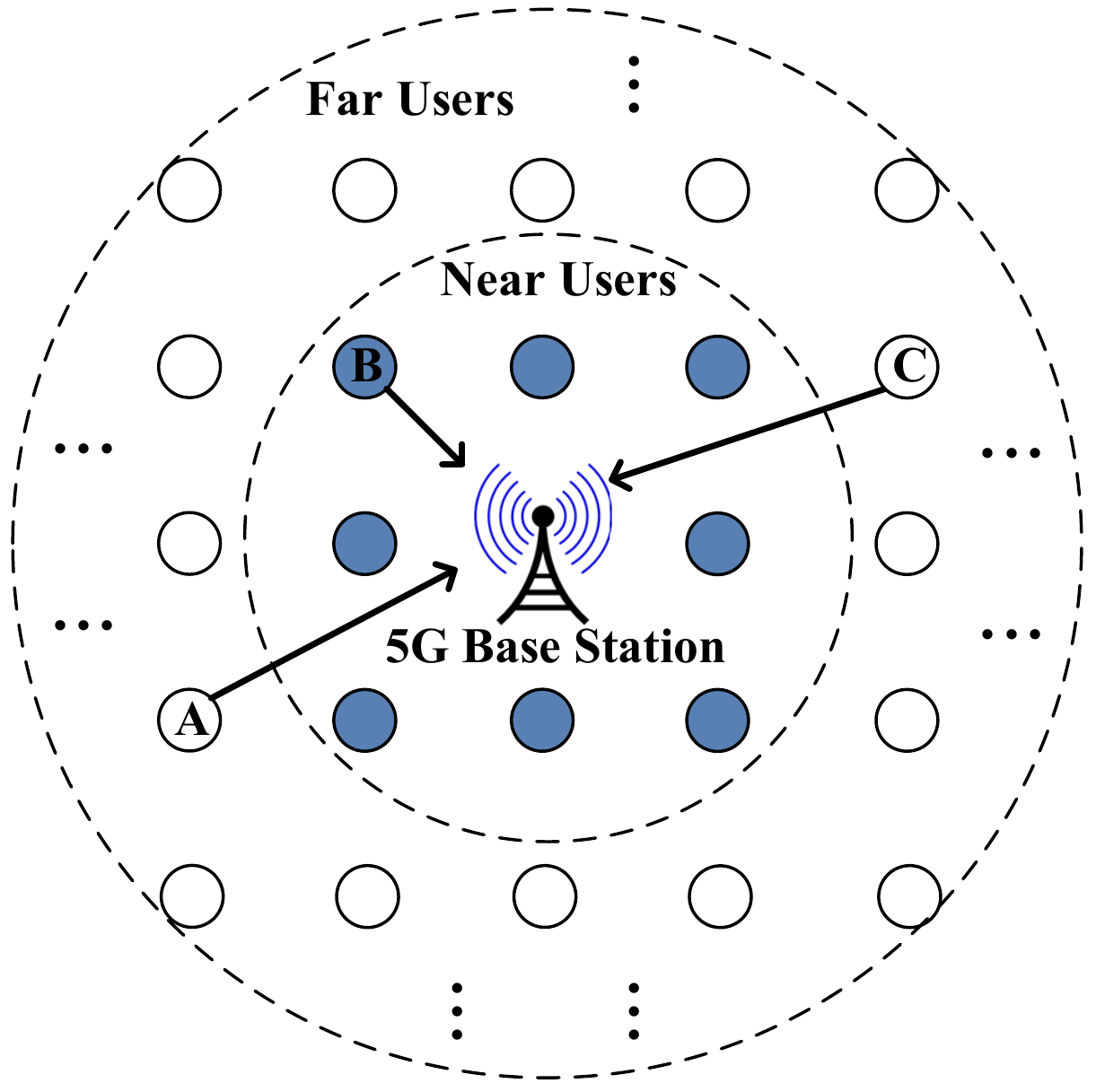}
\caption{An example of a 5G non-orthogonal multiple access (NOMA) system with one central base station (BS) and several uniformly distributed end users. Nodes A and C are weak users far from the BS, and node B is a strong user near the BS.}\label{fig:system_model}
\end{figure}

However, guaranteeing large received power differences within one SIC group is not always possible, especially when there is a disparity between the number of weak users and the number of strong users. This scenario is common in practical systems. An example is when users are uniformly distributed geographically around a BS, as shown in Fig. \ref{fig:system_model}. The peripheral area is larger than the area near the BS. Thus, there are more weak users than strong users. SIC-based NOMA may not work well in this scenario. A NOMA scheme in which weak users can also be grouped together, even though their received powers at the BS are nearly balanced, is highly desirable in practice. This paper is an attempt to fill this gap.

Our investigation of power-balanced NOMA leads us to the following two key design decisions:
\begin{itemize}\leftmargin=0in
\item [(1)] \textbf{\underline{Strong User Operations}}: When there are more strong users than weak users in a NOMA system, information-theoretically, grouping strong users together can only give small rate gain over conventional OMA schemes (e.g., TDMA), as will be detailed in Section \ref{sec:shortcommingSIC}. We simply opt to use TDMA for the ``excess'' strong users $-$ specifically, in each NOMA group, we pair one strong user with one weak user; the excess ungrouped strong users will adopt TDMA.
\item [(2)] \textbf{\underline{Weak User Operations}}: When there are more weak users than strong users, we may group more than one weak user into the same NOMA group. This is because, information-theoretically, grouping weak users can lead to large rate gain. However, this rate gain cannot be easily realized using conventional SIC. We put forth a new practical NOMA scheme, named Network-Coded Multiple Access (NCMA), to achieve the rate gain and to boost throughputs.
\end{itemize}

In this paper, we focus on scenario (2) (as argued previously, this scenario is more likely than the more-strong-users-than-weak-users scenario (1)). Unlike all the previous NOMA studies in the literature that focused on the use of MUD/SIC only, NCMA jointly exploits physical-layer network coding (PNC) and MUD to boost throughput of multipacket reception systems. PNC, first proposed in 2006, is a technique that turns mutual interference between signals from simultaneously transmitting users into useful network-coded information \referred{PNC06}\cite{PNC06}. Unlike SIC, a distinguishing feature of PNC is that it performs well when different users' received powers are balanced or near balanced. Experiments in \referred{NCMA1}\cite{NCMA1} showed that in a two-user NCMA system, when the MUD/SIC decoder failed to decode the native packets from two simultaneously transmitting users, with probability 40\%, a PNC decoder can still decode a network-coded packet at SNR of 8.5dB in a software-defined radio prototype under near power-balanced scenarios. A subtlety is that such decoded PHY-layer PNC packets are not useful for NOMA directly since NOMA aims for the native packets rather than the network-coded packet. A salient feature of NCMA is that it makes use of another layer of MAC channel coding to introduce correlations among PHY-layer packets in such a way that the PHY-layer PNC packets can be used to improve the overall NOMA throughput (the operation principles of a two-user NCMA system are reviewed in Section \ref{sec:NCMAoverview2}).

This paper considers a comprehensive design for multiuser NCMA targeted for 5G systems (previous NCMA work considered two users only). For multiuser NCMA, the decoding complexity increases exponentially with the number of users. To reduce complexity, we cluster active users into different groups (like SIC) and limit the group size. Thanks to PNC decoding, NCMA user grouping does not require large power differences between simultaneously received signals. Therefore, weak users can be grouped together. In particular, as will be seen in Section \ref{sec:NCMAoverview5}, experimental results show that with NCMA, grouping weak users together and allowing them to transmit for a longer period of time can substantially (i.e., share their allocated transmission times) improve the weak users' throughputs substantially.

Another shortcoming of previous NCMA systems is that they require all users in a group to use the same signal modulation \referred{NCMA1,NCMA2,MIMONCMA_Globecom}\cite{NCMA1,NCMA2,MIMONCMA_Globecom}. We refer to these systems as \emph{rate-homogeneous} NCMA. While rate-homogeneous NCMA can achieve substantial throughput improvement over SIC-based NOMA, it does not fully exploit weak users' channel conditions under the near power-balanced scenario (in practical systems, the weak users may still have slightly varying SNRs). In Section \ref{sec:NCMAmodulation1}, we show that forcing all users to use the same rate (modulation) may prevent the higher-SNR users from fully exploiting their superior channel conditions. In particular, the users with poor uplink channel conditions become the bottleneck of the whole group. To better exploit different channel conditions, this paper considers the use of different modulations for different weak users. Such systems are referred to as \emph{rate-diverse} NCMA. In particular, we put forth a symbol-splitting channel coding and modulation scheme, referred to as \emph{symbol-splitting encoding}, to enable channel-coded rate-diverse PNC decoding. Our experiments show that with symbol-splitting encoding, rate-diverse NCMA outperforms rate-homogeneous NCMA systems by around 80\% in terms of total system throughput.

To sum up, we have three major contributions:
\begin{itemize}\leftmargin=0in
\item [(1)] We are the first to study multiuser NOMA systems with power-balanced (near power-balanced) users. We show that a rate-homogeneous multi-user NCMA system design can substantially improve system throughput (over SIC-based NOMA) when weak users with near-balanced powers are grouped together.
\item [(2)] We further put forth a rate-diverse NCMA system design that further improves the NOMA system throughput (over the rate-homogeneous NCMA system). In particular, we provide the first design of a channel-coded rate-diverse PNC decoder, which is a key component in the overall rate-diverse NCMA system.
\item [(3)] We demonstrate the practical feasibility of multiuser rate-diverse NCMA via a real implementation on a software-defined radio platform. Our experimental results show that rate-diverse NCMA can boost the throughput of power-balanced NOMA in a practical setting.
\end{itemize}

The rest of the paper is organized as follows. We first overview related literature in Section \ref{sec:relatedwork}. Theoretical rate gain of power-balanced NOMA and the state-of-the-art SIC-based NOMA system design are re-examined rigorously in Section \ref{sec:shortcommingSIC}. We then give an overview of the NCMA system (a practical solution for power-balanced NOMA) in Section \ref{sec:NCMAoverview}. The modulation and channel coding designs of our proposed multiuser rate-diverse NCMA are studied in Sections \ref{sec:NCMAmodulation} and \ref{sec:ratediverseNCMA}. Section \ref{sec:experiments} presents the experimental results, and Section \ref{sec:conclusion} concludes the paper.

\section{Related Work}\label{sec:relatedwork}
\subsection{5G Non-Orthogonal Multiple Access (NOMA)}\label{sec:relatedwork1}
NOMA is a key enabler for the next generation communication systems to accommodate a large number of devices. Several NOMA schemes have been proposed for 5G. To allow a large number of users to share the same resource block, spreading codes \referred{MUSA}\cite{MUSA}, structured coding matrices \referred{SCMA,PDMA}\cite{SCMA,PDMA} and interleavers \referred{IDMA}\cite{IDMA} were used. Many NOMA studies also focused on successive interference cancellation (SIC) and different channel conditions between the users for multiplexing (e.g., SIC requires large power differences between users in the same group)\referred{powerdomainNOMA}\cite{powerdomainNOMA}. We refer to such systems as SIC-based NOMA in this paper.

SIC-based NOMA clusters users into different groups and tries to maximize the differences of the received signal powers in each group so that SIC can work well \referred{NOMAdynamicgroup}\cite{NOMAdynamicgroup}. However, grouping users with near balanced powers is inevitable in practical systems when we cannot ensure there are comparable numbers of strong and weak users. Our paper here puts forth a scheme for power-balanced (near power-balanced) NOMA as a complement to power-imbalanced NOMA.

\subsection{Physical-layer Network Coding for Multiple Access}\label{sec:relatedwork2}
PNC was originally proposed to increase the throughput of a two-way relay network (TWRN). It can double the throughput of a TWRN compared with the conventional store-and-forward relaying scheme \referred{PNC06}\cite{PNC06}. PNC has been studied and evaluated in depth during the past decade, and we refer the interested readers to \referred{liew2015primer,popovski2006anti,Nazer2011ReliablePNC}\cite{liew2015primer,popovski2006anti,Nazer2011ReliablePNC} and the references therein for details. Prior works on PNC focused almost exclusively on relay networks. By contrast, NCMA was the first attempt to apply PNC to non-relay networks (i.e., wireless multiple access networks) \referred{NCMA1,NCMA2,MIMONCMA_Globecom}\cite{NCMA1,NCMA2,MIMONCMA_Globecom}. We remark that all the previous NCMA work focuses on two users only. This paper is the first attempt in applying NCMA in multiuser NOMA system.

Besides NCMA, recently there have been other efforts to apply network coding (including PNC) in multiple access networks. For example, \referred{Cocco2011,CodedRandomAccess,SeekandDecode,jinhong2016PNC,RAwithPNC}\cite{Cocco2011,CodedRandomAccess,SeekandDecode,jinhong2016PNC,RAwithPNC} explored forming linear equations from the collided packets and derived source packets by solving the linear equations. However, \referred{Cocco2011,CodedRandomAccess}\cite{Cocco2011,CodedRandomAccess} only compute one equation for each overlapped packet, whereas NCMA can have more than one equation for each overlapped packet under favorable channel conditions. Furthermore, the decoding in \referred{RAwithPNC, SeekandDecode,jinhong2016PNC}\cite{RAwithPNC, SeekandDecode,jinhong2016PNC} is based on PHY-layer equations only, while NCMA makes use of an outer MAC-layer channel coding scheme to achieve better utilization of the PHY-layer PNC packets. Importantly, most existing works are theoretical in nature and lack implementation and experimental validations. They simply assume all users adopt the same signal modulations, even in fading channels. By contrast, our rate-diverse NCMA system takes into account the fact that different users are likely to experience different channel conditions under practical deployment scenarios (i.e., the near power balanced, but not exact power balanced scenario).

\subsection{Physical-layer Network Coding with Different Modulations}\label{sec:relatedwork3}
There have been some studies on PNC with different modulations in the literature. For example, \referred{HePNC,KoikeJSAC09}\cite{HePNC,KoikeJSAC09} considered non-channel-coded PNC schemes with different modulations. For reliable communication, channel-coded PNC is preferred \referred{liew2015primer}\cite{liew2015primer}. However, rate-diverse PNC decoder for channel-coded PNC systems has not been well studied. The schemes in \referred{HePNC,KoikeJSAC09}\cite{HePNC,KoikeJSAC09} are not applicable to channel-coded PNC because they do not preserve the linearity of the underlying channel codes. This paper puts forth a symbol-splitting encoding scheme that preserves channel-code linearity when different users adopt different modulations, thereby enabling reliable rate-diverse channel-coded PNC. As far as we know, this is the first rate-diverse channel-coded PNC design.

\section{Rate Gain in Power-Balanced NOMA and the Shortcoming of SIC}\label{sec:shortcommingSIC}
This section first presents the information-theoretical NOMA rate gain over conventional orthogonal multiple access (OMA) schemes. We argue that with equal powers, grouping strong users for NOMA does not give much rate gain over OMA, but grouping weak users give a large rate gain. Then, we present the shortcoming of SIC and argue that SIC-based NOMA may not be able to realize the weak users' potential rate gain. After that, we put forth our design strategies for power-balanced NOMA.

\subsection{Theoretical Rate Gain in Two-user Power-Balanced NOMA}\label{sec:shortcommingSIC1}
Let us assume two users as an example to study the theoretical NOMA rate gain. We argue that allowing two strong users to transmit together does not give much rate gain over conventional OMA schemes (e.g., TDMA), but allowing two weak users to transmit together does. Let $P_A$ and $P_B$ be the received powers of two strong users, say user A and user B, at the BS. Assume ${P_A} = {P_B} = P$ for simplicity. Fundamentally, the best possible NOMA (MUD) sum rate is ${R_{NOMA}} = \log (1 + P + P)$ \referred{tse2005fundamentals}\cite{tse2005fundamentals}, where the noise variance is normalized to be 1. The percentage rate gain $\eta$ of NOMA over OMA (using TDMA as an example) is
\begin{align}
\eta  = \frac{{{R_{NOMA}} - {R_{TDMA}}}}{{{R_{TDMA}}}} = \frac{{\log (1 + 2P) - \log (1 + P)}}{{\log (1 + P)}},
\label{equ:rate_gain}
\end{align}
\noindent where the sum rate of TDMA using time-sharing is ${R_{TDMA}} = \log (1 + P)$. It is easy to show that the rate gain $\eta$ in (\ref{equ:rate_gain}) monotonically decreases as $P$ increases. That is, the higher the SNR, the lower the rate gain. For example, when $P$ is 40dB, $\eta$ is only 0.07 (i.e., less than 10\%).

On the other hand, we notice that $\eta$ can be as large as 0.3 when $P$ is 8.5dB from (\ref{equ:rate_gain}). In other words, allowing weak users to transmit together may lead to a high rate gain for NOMA. However, we next show that SIC-based NOMA does not work well if both users are weak. That is, although the potential rate gain is high, it cannot be easily realized using conventional SIC.

\subsection{The Shortcoming of SIC-based NOMA}\label{sec:shortcommingSIC2}
Successive interference cancellation (SIC) for NOMA has been widely studied due to its simplicity. In SIC-based NOMA, different users are divided into small groups in such a way that within each group, the differences between the users' received powers at the BS (e.g., received SNRs) are maximized \referred{NOMAdynamicgroup}\cite{NOMAdynamicgroup}. However, ensuring large power differences within all SIC groups is not always possible. This is the case, for example, in practical scenarios in which the number of weak users and the number of strong users are unbalanced (see Fig. \ref{fig:system_model}).

\begin{figure}
\centering
\includegraphics[width=0.8\textwidth]{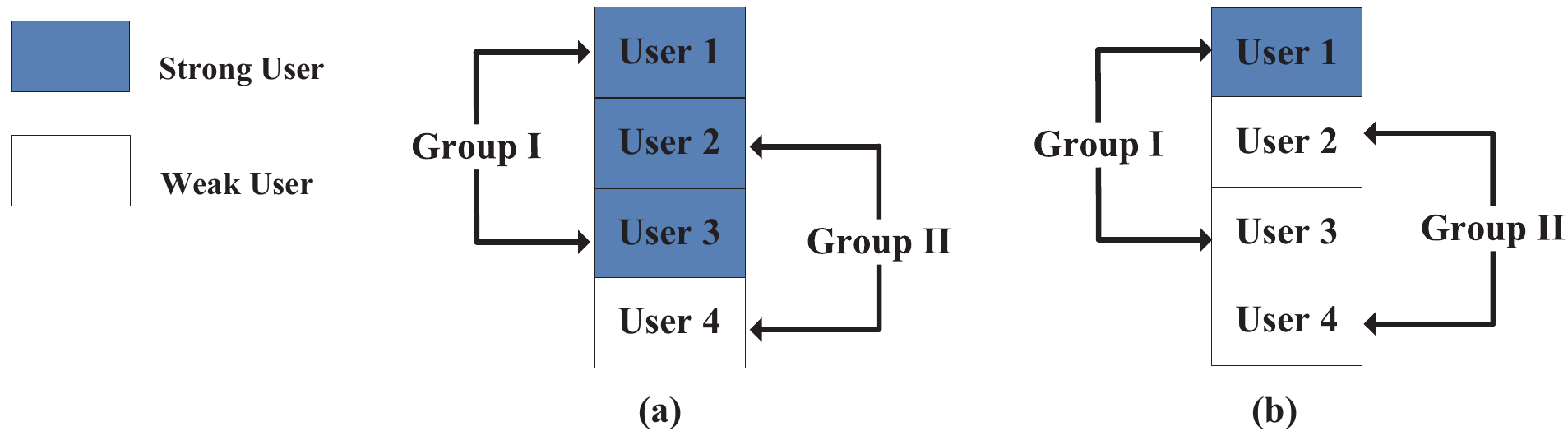}
\caption{Two examples of user grouping in SIC-based NOMA with four power-imbalanced users: (a) three strong users + one weak user, and (b) one strong user + three weak users. Users are ordered based on SNR, with user 1 having the largest SNR, and user 4 having the smallest SNR.}
\label{fig:sic_grouping}
\vspace{-0.12in}
\end{figure}

Here, let us look at the two specific examples shown in Fig. \ref{fig:sic_grouping}: in (a), we have three strong users + one weak user; in (b), we have one strong user + three weak users. In Fig. \ref{fig:sic_grouping}, users are ordered according to SNR, with user 1 having the largest SNR, and user 4 having the smallest SNR. Suppose that we want to divide the four users into two groups of two users each. However, for the scenario in Fig. \ref{fig:sic_grouping}(a), two strong users are inevitably grouped together, and in Fig. \ref{fig:sic_grouping}(b), two weak users are inevitably grouped together \referred{NOMAdynamicgroup}\cite{NOMAdynamicgroup}.

We now show that with balanced powers, the SIC decoder does not work well under the large inter-user interferences. Suppose that the two weak users, users 2 and 4 in Fig. \ref{fig:sic_grouping}(b) transmit packets $A$ and $B$ simultaneously, and assume that their powers $P_2$ and $P_4$ are equal (${P_2} = {P_4} = P$) at the BS for simplicity (note: the following analysis also applies to the two strong users, users 1 and 3 in Fig. \ref{fig:sic_grouping}(a)). To decode packet $A$ first, packet $B$ has to be treated as noise under SIC decoding, and hence the effective SINR for packet $A$ (or user 2) is
\begin{align}
{\rm{SINR}}_A = \frac{P}{{P + {\sigma ^2}}} < 0~dB,
\label{equ:SINR}
\end{align}

\noindent where ${\sigma ^2}$ is the noise power. Experimental results in \referred{NCMA1}\cite{NCMA1} showed that at SNR=8.5dB, with probability around 55\% the SIC decoder cannot decode any of the two packets for the power-balanced case. To realize the potential of grouping weak users with roughly equal powers in NOMA, alternative schemes other than SIC need to be used.

\subsection{Design Strategies for Power-balanced NOMA}\label{sec:shortcommingSIC3}
Based on the above discussions on the theoretical NOMA rate gain and the limitation of SIC, our investigation of power-balanced NOMA includes the following two key design strategies:

\textbf{(a) Strong User Operations:} As discussed in Section \ref{sec:shortcommingSIC1}, grouping two strong users leads to small rate gain over conventional OMA schemes. In this paper, we simply opt to use TDMA for strong users that are not grouped with weak users\footnote{From a practical viewpoint, grouping strong users with SIC also leads to high computational complexity. For instance, strong users usually adopt (1) high-order modulations (e.g., 64-QAM and beyond), and (2) advanced channel codes (e.g., LDPC codes). When two high-order modulated packets are superimposed, MUD/SIC may have to deal with: (1) a highly dense constellation map, and (2) an iterative channel decoder and the latencies therein \referred{liew2015primer}\cite{liew2015primer}.} (the excess strong users that cannot be paired with weak users, e.g., users 1 and 3 in Fig. \ref{fig:sic_grouping}(a)).

\textbf{(b) Weak Users Operations:} In Section \ref{sec:NCMAoverview}, we put forth a scheme, referred to as Network-Coded Multiple Access (NCMA), to realize the potential rate gain of grouping weak users\footnote{Besides the rate gain, grouping weak users also save more communication resources (i.e., transmission time) since weak users require more time to transmit the same amount of data  than the strong users.}. NCMA jointly exploits physical-layer network coding (PNC) and MUD to boost multiuser NOMA throughput. PNC tries to decode a network-coded packet (e.g., a bit-wise XOR packet $A \oplus B$ \referred{PNC06}\cite{PNC06}) rather than native packets $A$ and $B$. In particular, PNC does not require power imbalance to work well. \referred{NCMA1}\cite{NCMA1} shows that at 8.5dB SNR, when neither packet $A$ nor packet $B$ is decoded, with probability 40\% the PNC decoder can still decode $A \oplus B$. Although such PNC packets are not useful at the PHY layer, NCMA applies another layer of MAC channel coding to introduce correlations among PHY-layer packets such that these PNC packets can be useful at the MAC layer (see Section \ref{sec:NCMAoverview2} for details).

In addition to PNC decoding, NCMA uses a Reduced-constellation MUD (RMUD) decoder for native packet decoding when there are multiple weak users in the same group (see Section \ref{sec:ratediverseNCMA} for details). Although advanced iterative MUD/SIC decoding schemes \referred{Verdubook}\cite{Verdubook} that jointly decode multiple users are possible, such iterative schemes lead to high complexity and large latency. Since NCMA targets for real-time processing, we opt for the low-complexity non-iterative RMUD decoder.

Compared with previous studies of NCMA with coordinated access and two users only, this paper considers a comprehensive design for multiuser NCMA targeted for 5G systems. Section \ref{sec:NCMAoverview} overviews the key concepts and gives further examples to illustrate the merits of NCMA. We show via experiments that NCMA can substantially improve the throughputs of weak users compared with SIC-based NOMA

\section{NCMA Overview}\label{sec:NCMAoverview}
This section presents Network-Coded Multiple Access (NCMA). Sections \ref{sec:NCMAoverview1} and \ref{sec:NCMAoverview2} review the NCMA system model with two simultaneously transmitting users. Section \ref{sec:NCMAoverview3} considers three-user NCMA as an example to motivate the proposed power-balanced multiuser NCMA system design and Section \ref{sec:NCMAoverview4} discusses the complexity issue. Section \ref{sec:NCMAoverview5} compares the throughputs of SIC-based NOMA and NCMA given by real-network experiment results.

\subsection{NCMA Encoding Process}\label{sec:NCMAoverview1}
\begin{figure*}[t]
   \begin{minipage}{0.42\linewidth}
     \centering
     \includegraphics[width=0.85\textwidth]{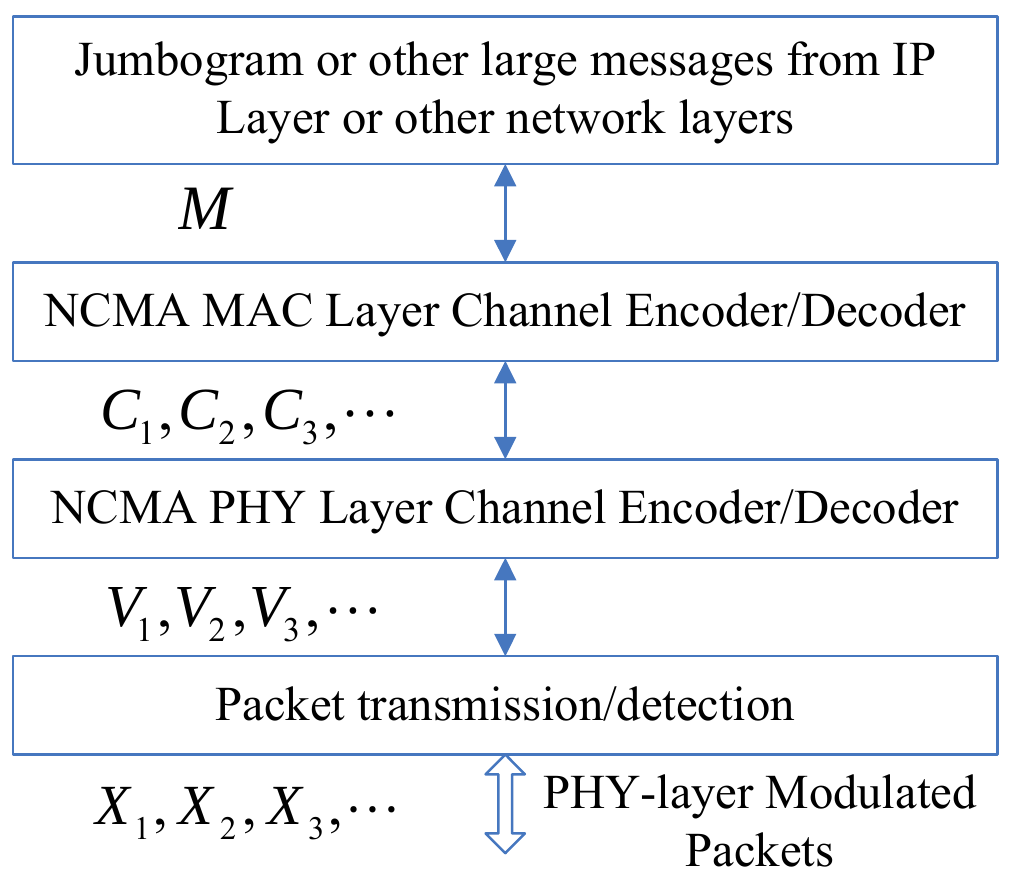}
     \caption{Encoding/decoding process for NCMA end users.}
     \label{fig:general_architec}
   \end{minipage}
   \hfill
   \begin{minipage}{0.56\linewidth}
      \centering
      \includegraphics[width=1.04\textwidth]{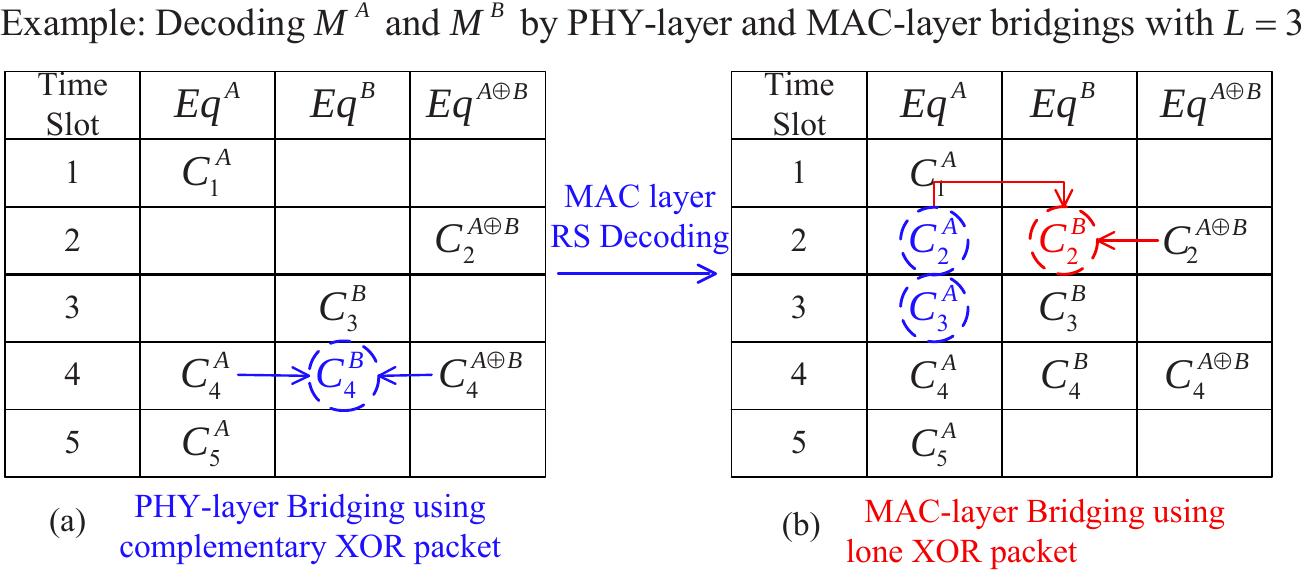}
      \caption{Two-user NCMA PHY-layer and MAC-layer bridgings example, using $L=3$ RS code.}
      \label{fig:two_user_example}
   \end{minipage}
   \vspace{-0.1in}
\end{figure*}


NCMA uses physical-layer network coding (PNC) and multiuser decoding (MUD) jointly to boost NOMA throughput. In \referred{MIMONCMA_Globecom}\cite{MIMONCMA_Globecom}, the NCMA receiver uses multiple antennas to accommodate high-order modulations beyond BPSK. The multiple-antenna NCMA is referred to as MIMO-NCMA. This paper assumes NCMA with two antennas at the BS unless otherwise specified.

NCMA includes both MAC-layer and PHY-layer operations. With respect to Fig. \ref{fig:general_architec}, at the MAC layer, a large message $M^s$ of user $s$, $s \in \Theta  = \{ A,B,C,...\}$, is divided and encoded into multiple packets, $C_i^s$, $i=1,2,$... . For simplicity, let us assume the use of Reed-Solomon (RS) code at the MAC layer when encoding a large message into multiple packets (other codes are also possible \referred{shenghaoFountaincode}\cite{shenghaoFountaincode}). At the PHY layer, each packet $C_i^s$ is further channel-encoded into $V_i^s$, and then modulated into $X_i^s$ for transmission. We adopt the convolutional code as the PHY-layer channel codes (other codes are also possible). Throughout the whole paper, we focus on a time-slotted NCMA system\footnote{The general idea of NCMA can also be applied to carrier sense multiple access (CSMA) systems or time-division multiple access (TMDA) systems by modifying MAC protocols to allow simultaneous transmissions by users.}\referred{NCMA1}\cite{NCMA1}. In this system, each user $s$ transmits packets $X_1^s,X_2^s,...,X_i^s$  to the BS in successive time slots. Packets of different end users can be configured to be transmitted in the same time slot.

\subsection{Review of Two-user NCMA}\label{sec:NCMAoverview2}

Let us briefly review the two-user NCMA system, and see how PNC and MUD can be jointly exploited to improve system throughput. In the uplink phase (note: NCMA focuses on the uplink transmissions from end users to the BS), users A and B transmit simultaneously. The BS then decodes the superimposed signals using two multiuser decoders at the PHY layer: the MUD decoder and the PNC decoder. The MUD decoder attempts to decode both packets $C_i^A$ and $C_i^B$ explicitly, and the PNC decoder attempts to decode\footnote{This paper only considers the bit-wise eXclusive-OR (XOR) operation, $ \oplus $ ,  of $C_i^A$ and $C_i^B$. Generalization beyond the XOR network coding operation is possible. } $C_i^A \oplus C_i^B$. In each time slot $i$, in general a subset of the set $\{C_i^A,C_i^B,C_i^A \oplus C_i^B\}$  is successfully decoded. The successfully decoded PHY-layer packets in different times slots are then collected and passed to the MAC layer. With the MAC-layer RS code, the BS can recover the original messages $M^A$ and $M^B$ after collecting enough packets from the set ${\{ C_i^A,C_i^B,C_i^A \oplus C_i^B\} _{i = 1,2,...}}$.

We next illustrate the essence of NCMA with a simple example. Fig. \ref{fig:two_user_example} shows an example of the decoding outcomes of the PNC and MUD decoder in five consecutive time slots. In time slot 4, $C_4^A$ and $C_4^A \oplus C_4^B$ (abbreviated as $C_4^{A \oplus B}$) are decoded. In this case, the PNC packet $C_4^{A \oplus B}$ can be used to recover the missing packet $C_4^B$ at the PHY layer. This process, which leverages the complementary PNC XOR packet, is referred to as \emph{PHY-layer bridging}\referred{NCMA1}\cite{NCMA1}. However, PHY-layer bridging cannot be applied directly to time slot 2 because neither native packet $C_2^A$ nor $C_2^B$  is available, and only the XOR packet $C_2^{A \oplus B}$ is decoded. In NCMA, such ``lone'' PNC packets, although not useful at the PHY layer, can be useful for MAC-layer decoding. Let us assume that $L=3$ native PHY-layer packets of user A (B) are needed to recover $M^A$  ($M^B$) at the MAC layer. In Fig. \ref{fig:two_user_example}(b), the BS has recovered enough native packets $C_i^A, i = 1,4,5$, to decode $M^A$ with the help of the MAC-layer RS code by time slot 5. This means that native packets $C_2^A$ and $C_3^A$ can also be recovered from $M^A$ (conceptually, we could re-encode $M^A$ to get $C_2^A$ and $C_3^A$, but in practice, a more efficient procedure is available \referred{NCMA2}\cite{NCMA2}). Accordingly, the original ``lone'' PNC packet $C_2^{A \oplus B}$ can now be combined with $C_2^A$ to recover $C_2^B$. Consequently, the BS now also has enough native packets (i.e., $L=3$) to recover the message of user B, $M^B$. We refer to this process as \emph{MAC-layer bridging}\referred{NCMA1}\cite{NCMA1}.

\subsection{Three-user NCMA }\label{sec:NCMAoverview3}
\begin{figure}
\centering
\includegraphics[width=0.97\textwidth]{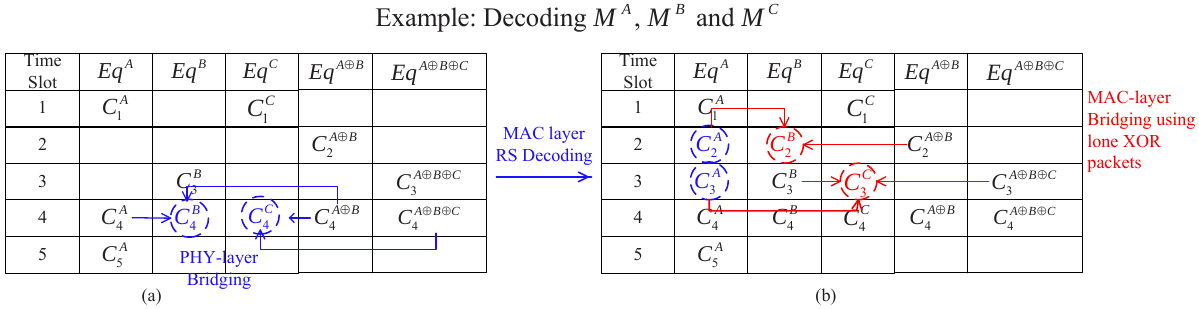}
\caption{Three-user NCMA PHY-layer and MAC-layer bridging example, generalized from Fig. \ref{fig:two_user_example}: (a) PHY-layer bridging; (b) MAC-layer RS decoding and bridging.}
\label{fig:three_user_example}
\vspace{-0.15in}
\end{figure}

Previous NCMA works \referred{NCMA1,NCMA2,MIMONCMA_Globecom}\cite{NCMA1,NCMA2,MIMONCMA_Globecom} were limited to two-user grouping only. In future 5G systems that support many users, allowing more NOMA concurrent transmissions can increase spectral efficiency. This subsection presents NCMA with three users as an example. We show that the underlying PHY-layer and MAC-layer decoder designs and bridging principles remain valid.

Suppose that users A, B, and C transmit their packets simultaneously in time slot $i$, and the BS receives their superimposed signals. At the PHY layer, three MUD decoders are needed for the BS to decode native packets $C_i^A$, $C_i^B$ and  $C_i^C$. Similarly, four PNC decoders are needed to get the four network-coded combinations $C_i^A \oplus C_i^B$, $C_i^A \oplus C_i^C$, $C_i^B \oplus C_i^C$ and $C_i^A \oplus C_i^B \oplus C_i^C$. That is, each PHY-layer decoder's output can be treated as a linear combination $aC_i^A \oplus bC_i^B \oplus cC_i^C$, where $a,b,c \in \{ 0,1\} $ and at least one of them must be 1. Fig. \ref{fig:three_user_example} shows an example of three-user NCMA by adding two more decoding outcome columns only, $C_i^C$ and $C_i^{A \oplus B \oplus C}$, to Fig. \ref{fig:two_user_example}.

It is worth emphasizing that there are more types of PHY-layer bridging for the three-user case than for the two-user case. For the three-user case, PHY-layer bridging can also happen between two PNC packets (namely, the XORed packets); for the two-user case, it can only happen between one PNC packet and one native packet. For instance, in time slot 4 of Fig. \ref{fig:three_user_example}(a), the missing individual packet $C_4^C$ can be recovered by XORing $C_4^{A \oplus B}$ and $C_4^{A \oplus B \oplus C}$. At the MAC layer, for the two-user case, all complementary PNC packets are resolved into native packets, and they do not need to be forwarded to the MAC layer. For the three-user case, a native packet and an ``unresolved'' packet can be forwarded to the MAC layer. For example, in time slot 3, the PNC packet $C_3^{A \oplus B \oplus C}$ is an ¡°unresolved¡± packet even though the BS has obtained $C_3^B$ at the PHY layer, since no PHY-layer bridging happens between $C_3^B$ and  $C_3^{A \oplus B \oplus C}$ (i.e., when we XOR  $C_3^B$ with $C_3^{A \oplus B \oplus C}$, we have $C_3^{A \oplus C}$ which is a PNC packet rather than a native packet).

\subsection{Complexity Issue in Multiuser NCMA System Design}\label{sec:NCMAoverview4}
Although the basic principles of two-user NCMA also apply to multiple users, the complexity issue arises in implementing a practical NCMA system with many active users in a group. First, the number of possible PHY-layer decoders increases exponentially with the number of users in a group ¨C e.g., there are ${2^N} - 1$ possible PNC and MUD decoders at the PHY layer where $N$ is the number of users in a group. Second, the complexity of MAC-layer decoding also increases with $N$. It is important for practical NCMA to have low-complexity PHY and MAC layer decoders while retaining good performance. In the experimental part of this paper, to contain complexity, we consider a maximum $N$ of 3. In addition, we only consider non-iterative decoders to ensure real-time performance in our software-defined radio prototype.

Besides the complexity associated with decoding, there is also the complexity associated with the identification of active users and their grouping therein. Specifically, before actual NCMA transmissions, the BS must identify active users who have packets to send. If the BS simply adopts a polling strategy, but many of transmitters are inactive with no packet to send (e.g., the non-saturated case), polling every user will be wasteful. After active user identification, the users are then divided into groups for simultaneous transmissions.

To contain the complexity associated with active user identification and grouping, we put forth a distributed reservation and grouping scheme that can identify active users and group them within a short time. We refer to this process as \emph{NCMA Random Access and Grouping procedure} (NCMA-RAG). NCMA-RAG improves the efficiency of channel access greatly compared with conventional polling schemes (details of NCMA-RAG can be found in Appendix \ref{sec:NCMA_RAG}).

Although conventional SIC-based NOMA also clusters users into groups to lower the SIC decoding complexity, the user grouping in NCMA-RAG differs in several aspects, thanks to PNC decoding in NCMA. NCMA user grouping does not require large power difference between simultaneously received signals. Therefore, weak users can be grouped together. By allowing multiple weak users to be grouped together and to share their time slots, each of the weak users can transmit for a longer period of time, thereby improving the weak users' throughputs. This solves the throughput degradation problem when there are more weak users than strong users in NOMA. In the rest of this paper, we focus on an NCMA group with weak users.

\begin{figure}
\centering
\includegraphics[width=0.47\textwidth]{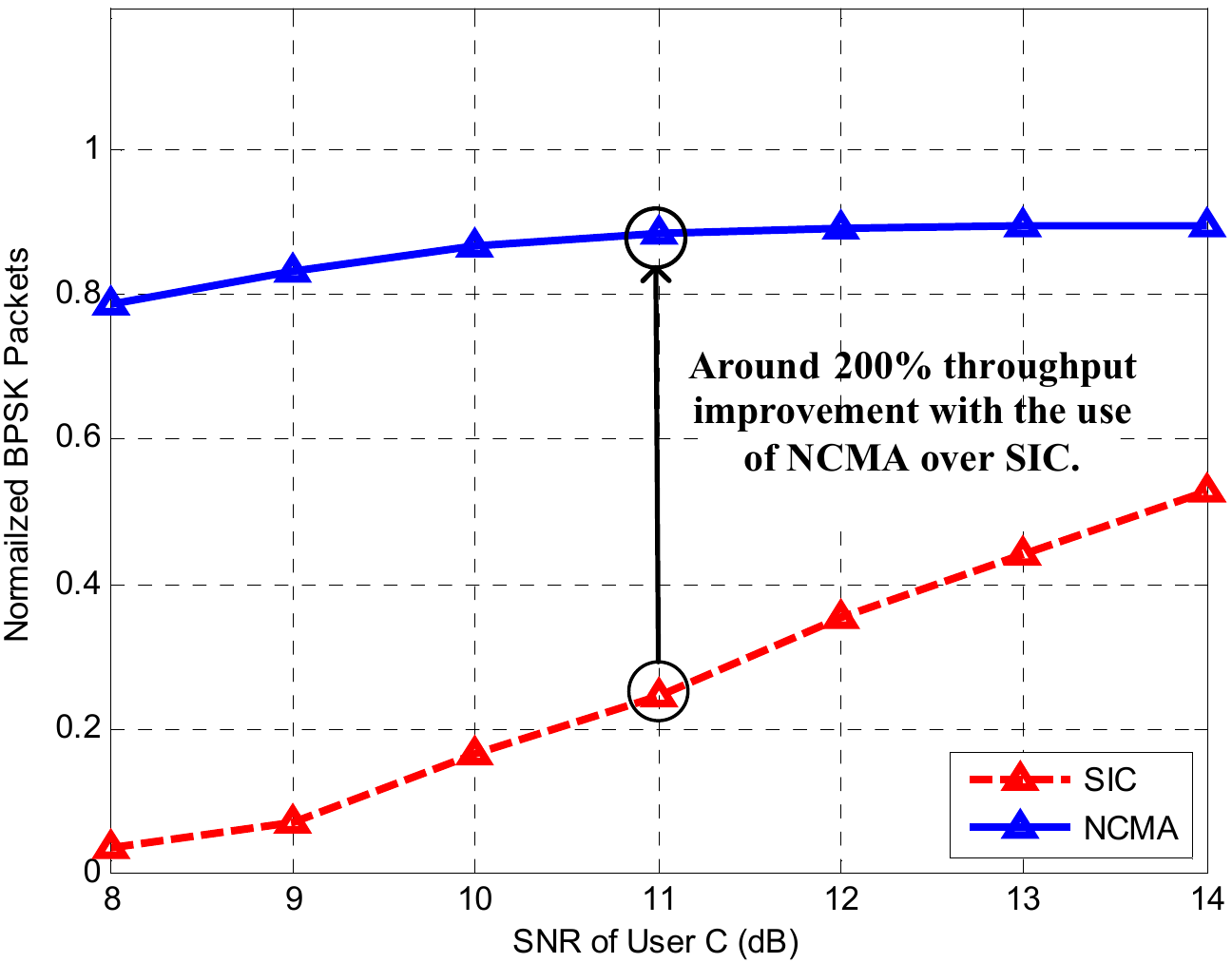}
\caption{Experimental results of user C's throughputs under SIC-based NOMA and NCMA-based NOMA. There are three weak users in the group: users A, B and C. The SNRs of users A and B are fixed at 8dB while the SNR of user C varies from 8dB to 14dB. All three users adopt BPSK modulation.}
\label{fig:throughput_sic_noma}
\vspace{-0.15in}
\end{figure}

\subsection{NCMA Works Well with Balanced Powers}\label{sec:NCMAoverview5}
Let us take a look at our experimental results of SIC-based NOMA and NCMA-based NOMA. The detailed experimental setup can be found in Section \ref{sec:experiments1}. We conducted experiments on our prototype with a group of three weak users A, B and C. All three users adopt BPSK modulation, and all three users send a packet in each time slot.  The throughput statistics were gathered over a large number of time slots. The SNRs of user A and user B were fixed at 8dB while the SNR of user C were varied from 8dB to 14dB.  Fig. \ref{fig:throughput_sic_noma} compares the numbers of successfully decoded packets per time slot for user C under SIC-based NOMA and NCMA-based NOMA.  For SIC-based NOMA, the SIC decoder was used. User C was decoded first by treating user A and user B as noise. For NCMA, both RMUD and PNC decoders were used, incorporating both PHY-layer and MAC-layer bridgings (details of RMUD and PNC decoders can be found in Section \ref{sec:ratediverseNCMA3}).

From Fig. \ref{fig:throughput_sic_noma}, we can see SIC-based NOMA gives a low throughput for user C. Although increasing user C's SNR can raise the throughput, the maximum throughput is still around 0.5 BPSK packet per time slot at 14dB SNR. The large inter-user interference causes the low throughput of user C. That is, in the weak users' group, we cannot treat user A and user B's signals as noise because the differences between their SNRs and user C's SNR are not large enough. Note that since user C's packet cannot be decoded in most cases, the throughputs of user A and user B will be even lower in SIC-based NOMA because the decoding of user A and user B depends on user C's packets being decoded correctly. For NCMA, we can see from Fig. \ref{fig:throughput_sic_noma} that user C's throughput can be increased much compared with SIC-based NOMA under all SNRs, e.g., around 200\% at 11dB SNR. The large throughput improvement is attributed to the joint use of RMUD and PNC. We also see that user C's throughput is bounded by one BPSK packet per time slot (the maximum possible normalized throughput in this setup).

Having demonstrated the superiority of NCMA over SIC-based NOMA under the above scenario, the next section explores whether NCMA can exploit user C's higher SNR to further raise the total throughput of the system beyond that demonstrated above.

\section{NCMA Modulation Design Issues}\label{sec:NCMAmodulation}
In practical systems, the channel conditions among weak users in the same group may vary (e.g., their SNRs can vary from 5dB to 15dB). The previous work on NCMA required all end users to adopt the same modulation order. We refer to this approach as \emph{rate-homogeneous} NCMA\footnote{Here, by ``rate'', we mean the modulation order that corresponds to the PHY-layer data rate, assuming different users use the same baud rate and channel code. More generally, a rate-diverse system can also be created if different users use the same modulation, the same baud rate, but different channel codes at different code rates; however, in this case, simple XOR-CD decoding for PNC cannot be applied because of the use of different channel codes \referred{liew2015primer}\cite{liew2015primer}.}. As shown in Fig. \ref{fig:throughput_sic_noma}, user C's throughput is upped-bounded by one BPSK packet when all the three users adopt BPSK. Section \ref{sec:NCMAmodulation1} further elaborates the low-throughput problem in rate-homogeneous NCMA. Section \ref{sec:NCMAmodulation2} then explore NCMA systems where users can use different modulations to better exploit their SNRs, referred to \emph{rate-diverse} NCMA. Specifically, we focus on BPSK and QPSK modulations for the weak users. We describe the subtle issues in PNC decoding in rate-diverse NCMA.

\subsection{Low-throughput Problem in Rate-homogeneous NCMA}\label{sec:NCMAmodulation1}
\begin{figure}
\centering
\includegraphics[width=0.8\textwidth]{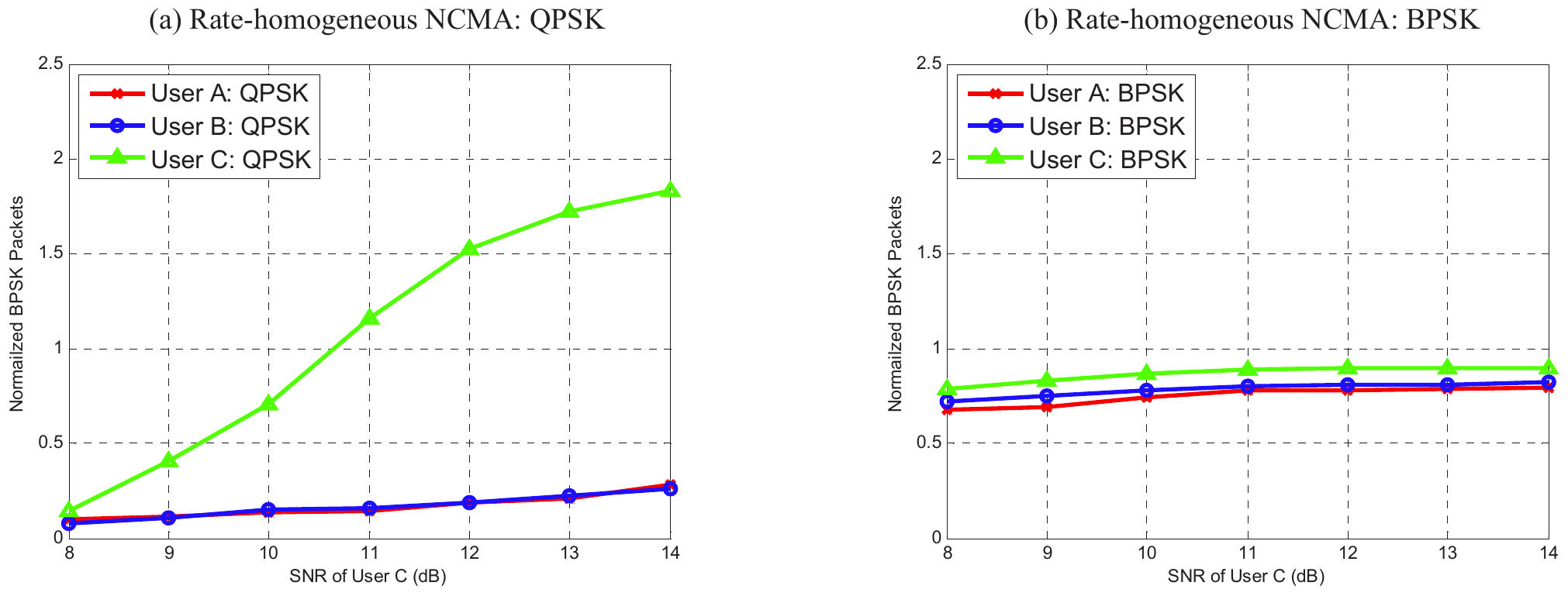}
\caption{Normalized experimental throughputs of three-user rate-homogeneous NCMA systems, assuming (a) QPSK and (b) BPSK. The y-axis stands for the normalized number of BPSK packets per time slot at the PHY layer, and the x-axis is the SNR of user C. Both users A and B's SNRs are set to be 8dB, and user C's SNR varies from 8dB to 14dB. }
\label{fig:throughput_rate_homo}
\vspace{-0.1in}
\end{figure}
Fig. \ref{fig:throughput_rate_homo}(a) and Fig. \ref{fig:throughput_rate_homo}(b) show the experimental throughputs per time slot of individual users in rate-homogeneous NCMA systems, assuming QPSK and BPSK, respectively. The experimental setup is the same as in Fig. \ref{fig:throughput_sic_noma}. Specifically, users A and B's SNRs are fixed at 8dB, and user C's SNR varies from 8dB to 14dB. Also, note that in Fig. \ref{fig:throughput_rate_homo} we treat one QPSK packet as two BPSK packets for fair throughput comparison. We observe the following:
\begin{itemize}\leftmargin=0in
\item [(1)] In Fig. \ref{fig:throughput_rate_homo}(a), all users adopt QPSK. Both users A and B have low throughputs because of their low SNRs, and the modulation order is not commensurate with their SNR. The throughput of user C, however, approaches 2 as its SNR increases;
\item [(2)] In Fig. \ref{fig:throughput_rate_homo}(b), all users adopt BPSK. Both users A and B can have higher throughputs than in Fig. \ref{fig:throughput_rate_homo}(a). However, the throughput of user C is upper bounded by 1 and drops by around 100\% (i.e., from QPSK to BPSK). User C cannot leverage its higher SNR to obtain higher throughput because of its use of BPSK;
\item [(3)] In both cases, the total system throughput is below 3.
\end{itemize}

For a practical multiple access system, it is unlikely that all users' uplink SNRs at the BS are exactly the same, even within a weak users' group. Rate-homogeneous NCMA forces all users to use the same modulation by ignoring their individual channel conditions, and therefore the uplink with the poorest channel condition becomes the bottleneck of the whole group. We next ask a simple but fundamental question: \textbf{can NCMA allow different users to use different modulations; and if yes, how can the system throughput benefits by doing so?}

\subsection{PNC Decoding Problem in Rate-diverse NCMA}\label{sec:NCMAmodulation2}
To solve the low-throughput problem in rate-homogeneous NCMA, we put forth \emph{rate-diverse NCMA}, where different users adopt different modulation orders to better utilize the channel conditions. To accommodate rate diversity in NCMA, we need to address a critical issue: how can PNC mapping be performed under different modulations (e.g., the XOR operation between different modulated symbols), while maintaining the linearity of channel codes at the same time?

To support PHY-layer real-time processing, a non-iterative PNC decoder, called XOR-CD (XOR Channel Decoding) is used in the NCMA system in \referred{NCMA1,NCMA2,MIMONCMA_Globecom}\cite{NCMA1,NCMA2,MIMONCMA_Globecom}. We first explain how XOR-CD works when different users adopt the same modulation order and code rate (i.e., the rate-homogeneous case). After that, we explain the problem of XOR-CD with different modulations (i.e., the rate-diverse case). For simplicity, here we assume the BS has one receive antenna, and three users, user A, user B and user C, transmit packets $C^A$,  $C^B$ and $C^C$ to the BS, respectively. Extensions to multiple antennas can be found in Section \ref{sec:ratediverseNCMA3}.


\begin{figure}
\centering
\includegraphics[width=0.75\textwidth]{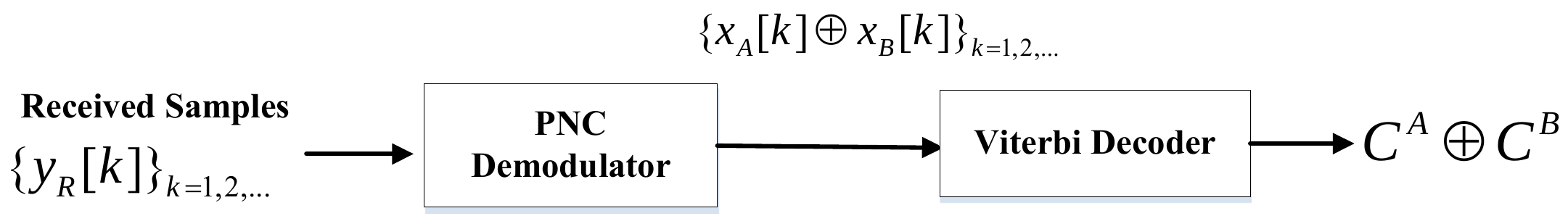}
\caption{NCMA PHY-layer PNC decoder (XOR-CD) Structure.}
\label{fig:xorcd}
\vspace{-0.15in}
\end{figure}

\subsubsection{PNC XOR operation for rate-homogeneous NCMA}
The general architecture for XOR-CD is shown in Fig. \ref{fig:xorcd}. We adopt the $[133, 171]_8$ rate-1/2 convolutional code. A salient feature of XOR-CD is that the standard point-to-point Viterbi channel decoder can be used directly without any changes to support real-time decoding \referred{NCMA2}\cite{NCMA2}. With respect to Fig. \ref{fig:general_architec}, ${V^s} = ({v_s}[1],...,{v_s}[n],...)$ is the PHY-layer codeword of user $s$ in one time slot (i.e., one binary convolutional-encoded packet of $C^s$), where ${v_s}[n] \in \{ 0,1\} $ is the $n$-th convolutional encoded bit. Assuming the modulation order is $m$ (e.g., $m=2$ for BPSK and $m=4$ for QPSK), the PHY-layer transmitted packet can be expressed as ${X^s} = ({x_s}[1],...,{x_s}[k],...)$, and ${x_s}[k]$ is the $k$-th modulated symbol of user $s$.

Let us assume an orthogonal frequency-division multiplexing (OFDM) system where multipath fading can be dealt with by cyclic prefix (CP). The $k$-th received sample in ${\{ {y_R}[k]\} _{k = 1,2,...}}$ in the frequency domain at the BS can be written as
\begin{align}
{y_R}[k] = {h_A}[k]{x_A}[k] + {h_B}[k]{x_B}[k] + {h_C}[k]{x_C}[k] + w[k],
\label{equ:receive_y}
\end{align}

\noindent where $w[k]$ are additive white Gaussian noises (AWGN) with variance ${\sigma ^2}$, and ${h_s}[k]$ is the channel gain of the $k$-th sample of user $s$.

Suppose that we want to decode the PNC packet $C^A \oplus C^B$. The received samples ${\{ {y_R}[k]\} _{k = 1,2,...}}$ are first passed through a PNC demodulator to obtain the XOR bits ${\{ {v_A}[n] \oplus {v_B}[n]\} _{n = 1,2,...}}$. The outputs from the PNC demodulator can be hard or soft bits. These XOR bits are fed to a standard Viterbi decoder (as used in a point-to-point system) to decode the network-coded packet $C^A \oplus C^B$. Since the two users A and B make use of the same code rate, the standard Viterbi decoder can be used because XOR-CD exploits the linearity of linear channel codes (note: convolutional codes are linear; XOR-CD will work with other linear codes as well). Specifically, define $\Pi ( \cdot )$ as the channel coding operation. Since  $\Pi ( \cdot )$ is linear, we have
\begin{align}
{V^A} \oplus {V^B} = \Pi \left( {{C^A}} \right) \oplus \Pi \left( {{C^B}} \right) = \Pi \left( {{C^A} \oplus {C^B}} \right).
\label{equ:linearity}
\end{align}

\begin{figure}
\centering
\includegraphics[width=0.8\textwidth]{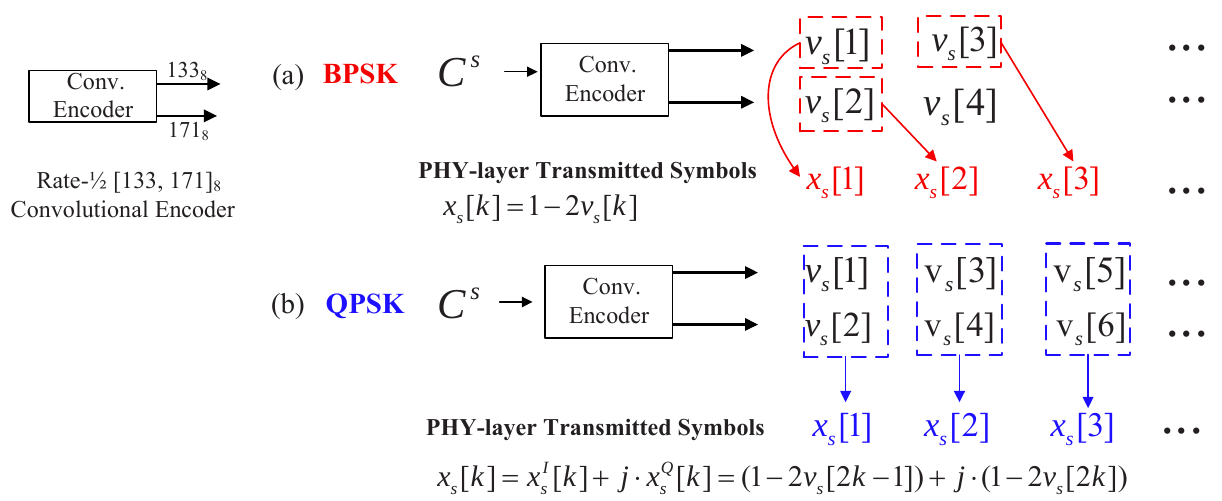}
\caption{Rate-1/2 $[133, 171]_8$ convolutional encoding and modulation procedure under: (a) BPSK, and (b) QPSK modulations. }
\label{fig:standard_coding}
\vspace{-0.15in}
\end{figure}

We give an example of how to obtain ${\{ {v_A}[n] \oplus {v_B}[n]\} _{n = 1,2,...}}$, assuming BPSK modulation for both users. The $k$-th BPSK modulated symbol ${x_s}[k]$ of the PHY-layer transmitted packet ${X^s}$ can be expressed as ${x_s}[k] = 1 - 2{v_s}[k]$ (see Fig. \ref{fig:standard_coding}(a)).

An important issue in PNC is how to calculate ${x_A}[k] \oplus {x_B}{\rm{[k]}}$ (abbreviated as ${x_{A \oplus B}}[k]$) using the received sample ${y_R}[k]$ in (\ref{equ:receive_y}), referred to as \emph{PNC mapping}. The BPSK PNC mapping for ${x_{A \oplus B}}[k]$ is defined as ${x_A}[k] \oplus {x_B}[k]={x_A}[k]{x_B}[k]$ given that ${x_A}[k],{x_B}[k] \in \{ 1, - 1\} $. The demodulation rule for the XORed bits is defined as
\begin{align}
{v_A}[k] \oplus {v_B}[k] = \frac{{1 - {x_A}[k]{x_B}[k]}}{2}.
\label{equ:bpsk_demo}
\end{align}

After that, ${\{ {v_A}[n] \oplus {v_B}[n]\} _{n = 1,2,...}}$ are fed to the Viterbi decoder to decode the PNC packet ${C^A} \oplus {C^B}$. When different users adopt the same modulation, XOR-CD works in a similar way for higher-order modulations beyond BPSK (e.g., QPSK and 16-QAM) after each modulated symbol is mapped to bits \referred{MIMONCMA_Globecom}\cite{MIMONCMA_Globecom}.

\subsubsection{Difficulty in PNC XOR operation in rate-diverse NCMA}
We now explain the difficulty of applying XOR-CD with different modulations. Assuming user C uses QPSK, the $k$-th modulated symbol ${x_C}[k]$, ${x_C}[k] \in \{ 1 + j,1 - j, - 1 + j, - 1 - j\}$, of the PHY-layer packet ${X^C}$ is
\begin{align}
{x_C}[k] = (1 - 2{v_C}[2k - 1]) + j \cdot (1 - 2{v_C}[2k]),~~ k = 1,2,...,n,...
\label{equ:qpsk_standard}
\end{align}

That is, the odd (even) bits of the convolutional-encoded packet $V^C$ are mapped to the in-phase (quadrature) components of ${x_C}[k]$ in QPSK, i.e., $x_C^I[k] = 1 - 2{v_C}[2k - 1]$ ($x_C^Q[k] = 1 - 2{v_C}[2k]$). Fig. \ref{fig:standard_coding} illustrates the differences between BPSK and QPSK modulations. Note that for both BPSK and QPSK, the odd bits and even bits of $V^s$ are generated from two different code generator polynomials (i.e., $133_8$ and $171_8$ in the IEEE Standard). Since each QPSK symbol contains two bits, one from each polynomial, while each BPSK symbol contains only one bit from one of the polynomials, how to perform the proper PNC mapping (XOR-CD) for the overlapping QPSK and BPSK symbols is an issue. It is difficult to find a proper PNC mapping between a BPSK symbol and a QPSK symbol (e.g., ${x_A}[k] \oplus {x_C}[k]$ and ${x_B}[k] \oplus {x_C}[k]$) that maintains the linearity of convolutional codes as (\ref{equ:linearity}). Therefore, conventional XOR-CD decoder does not work for different modulations.

Fortunately, as will be seen in the next section, we can redesign the channel coding and modulation scheme that can enable PNC among different modulations, and by doing so, the advantages of NCMA can be fully exploited.

\begin{figure}
\centering
\includegraphics[width=0.8\textwidth]{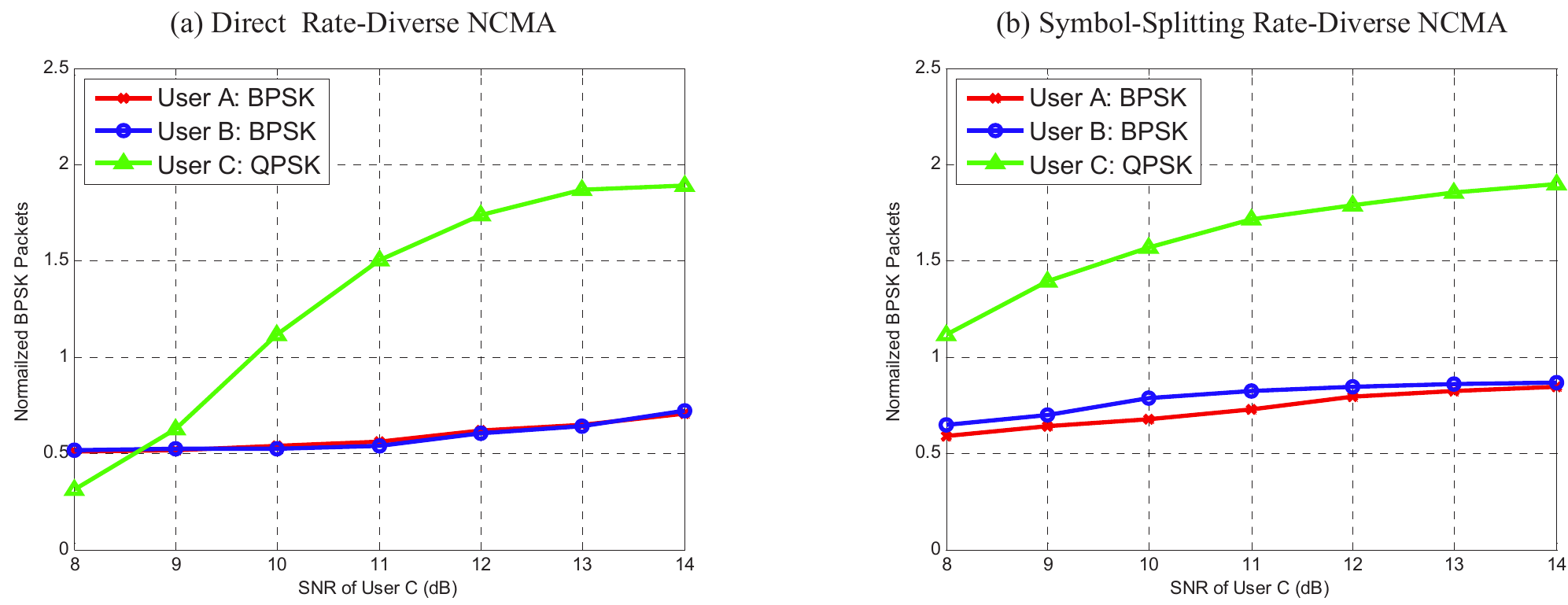}
\caption{Normalized experimental throughputs of three-user Rate-Diverse NCMA systems: (a) Direct Rate-Diverse NCMA and (b) Symbol-Splitting Rate-Diverse NCMA. We assume users A and B adopt BPSK, and user C adopts QPSK. The y-axis is the normalized number of BPSK packets per time slot at the PHY layer, and the x-axis is the SNR of user C. Both users A and B's SNRs are set to be 8dB, and user C's SNR varies from 8dB to 14dB. }
\label{fig:throughput_rate_diverse}
\vspace{-0.15in}
\end{figure}

\section{Rate-Diverse NCMA Modulation and Channel Coding Design}\label{sec:ratediverseNCMA}
This section presents rate-diverse NCMA that can fully exploit the varying SNRs among weak users. We study the case of a three-user NCMA group with two users, say users A and B, adopting BPSK, and one user, say user C, adopting QPSK (abbreviated as 2B1Q). We remark that the decoding principle for 2B1Q can be generalized to other scenarios easily, e.g., 2QPSK+1BPSK.

For 2B1Q, PNC decoding (XOR-CD) between the two BPSK users is the same as before (e.g., we can adopt the BPSK PNC mapping defined in Section \ref{sec:NCMAmodulation2}, and the calculations of each XORed bit's soft information will be presented in Section \ref{sec:ratediverseNCMA3}). But PNC decoding cannot happen between BPSK and QPSK users, and Section \ref{sec:ratediverseNCMA1} shows that this does not fully exploit the advantages of NCMA. Section \ref{sec:ratediverseNCMA2} presents our designs to enable PNC even among different modulations. Section \ref{sec:ratediverseNCMA3} presents the details of our rate-diverse NCMA PHY-layer decoders.

\subsection{Direct extension from rate-homogeneous NCMA}\label{sec:ratediverseNCMA1}
As discussed in Section \ref{sec:NCMAmodulation2}, PNC decoding does not work among different modulations. For 2B1Q, if we directly generalize rate-homogeneous NCMA to rate-diverse NCMA, only one possible PNC decoder is available (i.e., to decode $C^A \oplus C^B$). We refer to such an NCMA system as \emph{Direct Rate-diverse NCMA} (DR-NCMA).

Fig. \ref{fig:throughput_rate_diverse}(a) shows the experimental throughputs of individual users of DR-NCMA with the same setups as Fig. \ref{fig:throughput_sic_noma} and Fig. \ref{fig:throughput_rate_homo} (the details of experimental set-up can be found in Section \ref{sec:experiments1}), except that users A and B adopt BPSK, and user C adopts QPSK. PNC decoding is not applied to user C; it is applied to users A and B only. Therefore, user C does not participate in PHY-layer bridging or MAC-layer bridging. Compared with BPSK rate-homogeneous NCMA in Fig. \ref{fig:throughput_rate_homo}(b), the BPSK users have lower throughputs in DR-NCMA. Moreover, the performance of the QPSK user may still be suboptimal because only the MUD decoder is applied to it.

Overall, DR-NCMA does not fully exploit the advantages of NCMA, although the total system throughput is above 3 (when user C has high SNRs) and higher than those in the rate-homogeneous systems in Fig. \ref{fig:throughput_rate_homo}(a) and (b). Fortunately, we can enable PNC among different modulations by redesigning the standard channel coding and modulation scheme.

\subsection{Exploit PNC between Different Modulations}\label{sec:ratediverseNCMA2}
\begin{figure}
\centering
\includegraphics[width=0.8\textwidth]{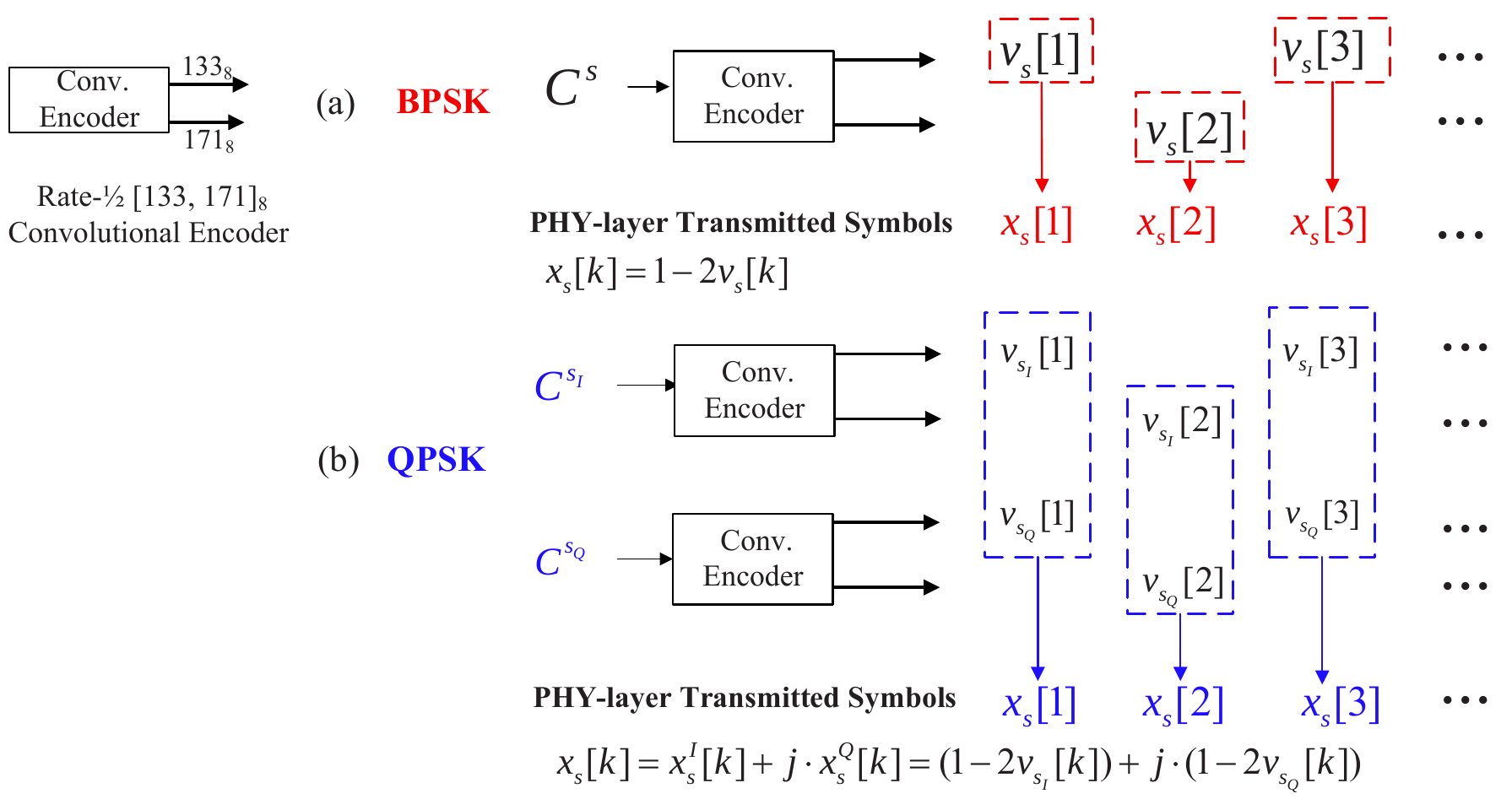}
\caption{Convolutional encoding and modulation schemes for Symbol-Splitting Rate-Diverse NCMA: (a) Same procedure for BPSK packets, and (b) Symbol-Splitting Encoding for QPSK packets. }
\label{fig:sp_coding}
\vspace{-0.15in}
\end{figure}

We now present our designs that enable PNC among different modulations. Let us focus on user A (BPSK) and user C (QPSK) as an example. We study how to perform PNC mapping between ${x_A}[k]$ and ${x_C}[k]$, where ${x_C}[k] = x_C^I[k] + j \cdot x_C^Q[k]$.

As discussed in Section \ref{sec:NCMAmodulation2}, an important design issue for PNC mapping is how to maintain the linearity of convolutional codes. Conventional channel encoding and modulation scheme fails to do so because within the overlapping QPSK and BPSK symbols,  $x_C^I[k]$ and $x_C^Q[k]$ are from two different polynomials, while ${x_A}[k]$ is from one of the two polynomials. However, if the in-phase and quadrature bits of the QPSK packet can be encoded separately as if they were from two BPSK packets (one containing in-phase bits; one containing quadrature bits), then the PNC mapping of ${x_A}[k]$ and $x_C^I[k]$, and the PNC mapping of ${x_A}[k]$ and $x_C^Q[k]$ become possible.

Fig. \ref{fig:sp_coding} presents our channel encoding and modulation scheme for QPSK in rate-diverse NCMA. For user C, let $C^{{C_I}}$ and $C^{{C_Q}}$ denote two small packets (which can be equally divided from $C^C$). They are separately convolutional encoded to ${V_{{C_I}}} = \{ {v_{{C_I}}}[1],{v_{{C_I}}}[2],...,{v_{{C_I}}}[n],...\}$ and ${V_{{C_Q}}} = \{ {v_{{C_Q}}}[1],{v_{{C_Q}}}[2],...,{v_{{C_Q}}}[n],...\}$, respectively. The $k$-th modulated symbol ${x_C}[k]$ for the QPSK packet ${X^C} = ({x_C}[1],...,{x_C}[k],...)$  is now defined as
\begin{align}
{x_C}[k] = x_C^I[k] &+ j \cdot x_C^Q[k] =  (1 - 2{v_{{C_I}}}[k]) + j \cdot (1 - 2{v_{{C_Q}}}[k]), k = 1,2,...,n,...
\label{equ:qpsk_sp}
\end{align}

That is,  $C^{{C_I}}$ ($C^{{C_Q}}$) is encoded to be the in-phase (quadrature) bits of the QPSK packet. We refer to this channel encoding and modulation scheme as \emph{symbol-splitting encoding}.

In essence, the symbol-splitting encoding scheme makes one QPSK packet equivalent to two ``small'' BPSK packets from the channel coding perspective, e.g., two BPSK packets are embedded in the in-phase and quadrature parts of one QPSK packet, respectively. Since each ``small'' BPSK packet is now encoded in the same way as a regular BPSK packet, we can define the PNC mapping between symbols ${x_A}[k]$ and ${x_C}[k]$ as\footnote{We can also define PNC mapping $x_C^I[k] \oplus x_C^Q[k]$ and compute the PNC packet ${C^{{C_I}}} \oplus C^{{C_Q}}$.  In symbol-splitting encoding, ${C^{{C_I}}}$ and ${C^{{C_Q}}}$ can be regarded as two packets with a fixed 90-degree relative phase offset (i.e., they will be encoded as the in-phase and quadrature parts of the QPSK packet). However, our experimental results show that ${C^{{C_I}}} \oplus C^{{C_Q}}$ does not give extra performance gain since MUD decoders that decode ${C^{{C_I}}}$ and ${C^{{C_Q}}}$ work well already. In this paper, we do not consider the PNC decoders that contain ${C^{{C_I}}} \oplus C^{{C_Q}}$. }
\begin{align}
x_{A \oplus C}^I[k] = {x_A}[k] \oplus x_C^I[k] = {x_A}[k] x_C^I[k], \notag \\
x_{A \oplus C}^Q[k] = {x_A}[k] \oplus x_C^Q[k] = {x_A}[k] x_C^Q[k].
\label{equ:rate_diverse_mapping}
\end{align}

With the same demodulation rule as in (\ref{equ:bpsk_demo}), the corresponding XOR bits ${\{ {v_A}[n] \oplus {v_{{C_I}}}[n]\} _{n = 1,2,...}}$ and ${\{ {v_A}[n] \oplus {v_{{C_Q}}}[n]\} _{n = 1,2,...}}$ obtained from the demodulator are then fed to the Viterbi decoder to decode ${C^A} \oplus C^{{C_I}}$ and ${C^A} \oplus C^{{C_Q}}$, respectively. That is, with symbol-splitting encoding, we can perform PNC decoding between BPSK and QPSK users.

\subsection{Symbol-Splitting Rate-Diverse NCMA }\label{sec:ratediverseNCMA3}
\begin{table}[t]
\centering
\caption{\textnormal{SR-NCMA PHY-layer Decoders, assuming three end users A, B and C. Users A and B use BPSK, and user C uses QPSK.}}
\begin{tabular}{|c|c|c|c|c|}
 \hline
 {MUD Decoder} & \multicolumn{2}{c|}{PNC Decoder} \\
 \hline
 ${C^A}$        & ${C^A} \oplus {C^B}$        & \\  \hline
 ${C^B}$         & ${C^A} \oplus C^{{C_I}}$       & ${C^A} \oplus C^{{C_Q}}$  \\  \hline
 $C^{{C_I}} $ & ${C^B} \oplus C^{{C_I}}$ &  ${C^B} \oplus C^{{C_Q}}$\\  \hline
 $C^{{C_Q}}$ &  ${C^A} \oplus {C^B} \oplus C^{{C_I}}$ & ${C^A} \oplus {C^B} \oplus C^{{C_Q}}$ \\  \hline
 \end{tabular}
\vspace{-0.15in}
 \label{tab:rate_diverse_decoder}
\end{table}

This subsection presents the rate-diverse NCMA system with the symbol-splitting encoding scheme for QPSK packets. We refer to this NCMA system as \emph{Symbol-splitting Rate-Diverse NCMA} (SR-NCMA). We first list different PHY-layer decoders used in SR-NCMA and compare the SR-NCMA system throughput with those of DR-NCMA and rate-homogeneous NCMA. After that, we present the details of PHY-layer decoders for SR-NCMA.

Section \ref{sec:ratediverseNCMA2} discussed two PNC decoders that decode ${C^A} \oplus C^{{C_I}}$ and ${C^A} \oplus C^{{C_Q}}$ between the BPSK user A and the QPSK user C. In general, with symbol-splitting encoding, there are total seven possible PNC decoders to decode different linear combinations between the three users A, B, and C, as shown in Table \ref{tab:rate_diverse_decoder}. Also, four MUD decoders can be used in SR-NCMA. In short, each PHY-layer decoder's output can be treated as a linear combination $aC^A \oplus bC^B \oplus cC^{{C_I}}$ or $aC^A \oplus bC^B \oplus cC^{{C_Q}}$, where $a,b,c \in \{ 0,1\} $ and at least one of them must be 1\footnote{The PHY-layer decoding complexity for SR-NCMA is acceptable. SR-NCMA and DR-NCMA amount to decode 11 and 5 equivalent BPSK packets (including MUD and PNC packets, and one QPSK packet is treated as two BPSK packets). The QPSK and BPSK rate-homogeneous NCMA amount to decode 14 and 7 BPSK packets, respectively. As will be seen in the Section \ref{sec:experiments2}, the total system throughput of SR-NCMA can be up to 80\% higher than that of DR-NCMA and rate-homogeneous NCMA.}.

Fig. \ref{fig:throughput_rate_diverse}(b) shows the experimental throughputs of individual users of SR-NCMA. Compared with DR-NCMA in Fig. \ref{fig:throughput_rate_diverse}(a), the BPSK users have higher throughputs in SR-NCMA, which is also comparable to BPSK rate-homogeneous NCMA in Fig. \ref{fig:throughput_rate_homo}(b). The throughput of the QPSK user also improves and converges to 2 quickly as SNR increases, thanks to the PNC packets between QPSK and BPSK users, e.g., the QPSK user can have PHY-layer and MAC-layer bridgings through PNC packets in SR-NCMA. For the total system throughput, SR-NCMA has the highest throughput compared with DR-NCMA and rate-homogeneous NCMA (e.g., approaches to 4 when user C has high SNRs). Overall, SR-NCMA allows users to select a proper modulation order to better utilize the channel conditions.

\textbf{Calculations of Soft Information in Demodulators:} We now explain how to obtain the soft information from the PHY-layer demodulators. For generality, we assume two antennas at the BS in this presentation. We focus on the soft information of ${\{ {v_A}[n] \oplus {v_{{C_I}}}[n]\} _{n = 1,2,...}}$ as an example, and other PNC and MUD demodulators can follow the same manner. The soft information of ${\{ {v_A}[n] \oplus {v_{{C_I}}}[n]\} _{n = 1,2,...}}$ are fed to the Viterbi decoder to decode packet ${C^A} \oplus C^{{C_I}}$.

Let the received frequency-domain samples on the two antennas at the BS be ${{\rm{\{ }}{y_{R1}}[k]{\rm{\} }}_{k = 1,2,3...}}$ and ${{\rm{\{ }}{y_{R2}}[k]{\rm{\} }}_{k = 1,2,3...}}$ (our NCMA system is an OFDM system). Our target is to compute the log-likelihood ratio (LLR) of ${v_A}[k] \oplus {v_{{C_I}}}[k]$, based on the $k$-th received samples ${y_{R1}}[k]$ and ${y_{R2}}[k]$ (in the following, we drop the index $k$ for simplicity):
\begin{align}
{y_{R1}} = {h_{A1}}{x_A} + {h_{B1}}{x_B} + {h_{C1}}{x_C} + {w_1}, \notag \\
{y_{R2}} = {h_{A2}}{x_A} + {h_{B2}}{x_B} + {h_{C2}}{x_C} + {w_2},
\label{equ:y1y2}
\end{align}

\noindent where ${h_{s1}}$ and ${h_{s2}}$ are the uplink channel gains of end user $s$ associated with the first and second antenna, respectively, and ${w_1}$ and ${w_2}$ are additive white Gaussian noises (AWGN) with variances $\sigma _1^2$ and $\sigma _2^2$. We assume the noise variances $\sigma _1^2$ and $\sigma _2^2$ to be the same. Note that, in real wireless systems, $\sigma _1^2$ and $\sigma _2^2$ may not be equal sometimes; however, our derivations below can be easily generalized to deal with the case $\sigma _1^2 \ne \sigma _2^2$.

Define the LLR of packet A's BPSK symbol (i.e., $x_A$) as $log(P_A/Q_A)$, where $P_A$ and $Q_A$ are the probabilities for $x_A$ to be 1 and -1, respectively. Similarly, for $LLR(x_A \oplus x_C^I)$,  $P_{A \oplus {C_I}}$ and $Q_{A \oplus {C_I}}$ are the probabilities corresponding to $x_A \oplus x_C^I = 1$ and $x_A \oplus x_C^I = -1$. We have
\begin{align}
LLR(x_A \oplus x_C^I) &= \log P_{A \oplus {C_I}}- \log Q_{A \oplus {C_I}} \notag\\
&= \log \Pr (x_A \oplus x_C^I = 1|{y_{R1}},{y_{R2}}) - \log \Pr (x_A\oplus x_C^I =  - 1|{y_{R1}},{y_{R2}}).
\label{equ:llr}
\end{align}

\newcounter{mytempeqncnt}
\begin{figure*}[!t]
\setcounter{mytempeqncnt}{\value{equation}}
\small
\begin{align}
\log &P_{A \oplus {C_I}}= \log \Pr (x_A \oplus x_C^I = 1|{y_{R1}},{y_{R2}}) \notag\\
\label{equ:logpnc_p1} &{\rm{     }} \propto \log \sum\limits_{({x_A},{x_B},{x_C}) \in {\chi _{{x_{A \oplus {C_I}}} = 1}}} {\exp \{  - \frac{{|{y_{R1}} - {h_{A1}}{x_A} - {h_{B1}}{x_B} - {h_{C1}}{x_C}{|^2}}}{{\sigma^2}}} {\rm{\} exp\{ }} - \frac{{|{y_{R2}} - {h_{A2}}{x_A} - {h_{B2}}{x_B} - {h_{C2}}{x_C}{|^2}}}{{\sigma^2}}{\rm{\} }},\\
\label{equ:logpnc_p1_sim} &{\rm{         }} \propto {\min _{({x_A},{x_B},{x_C}) \in {\chi _{{x_{A \oplus {C_I}}} = 1}}}}\{ |{y_{R1}} - {h_{A1}}{x_A} - {h_{B1}}{x_B} - {h_{C1}}{x_C}{|^2} + |{y_{R2}} - {h_{A2}}{x_A} - {h_{B2}}{x_B} - {h_{C2}}{x_C}{|^2}{\rm{\} }}.
\end{align}
\hrulefill

\end{figure*}

Out of the 16 constellation points associated with the symbols $({x_A},{x_B},{x_C})$, let ${\chi _{{x_{A \oplus {C_I}}} = 1}}$ denote the set of symbols $({x_A},{x_B},{x_C})$ that satisfy $x_A \oplus x_C^I = 1$.  We can express $\log P_{A \oplus {C_I}}$ as (\ref{equ:logpnc_p1}). $\log Q_{A \oplus {C_I}}$ can be computed in a similar way based on the set ${\chi _{{x_{A \oplus {C_I}}} =  - 1}}$. To further simplify (\ref{equ:logpnc_p1}), we adopt the log-max approximation, $\log (\sum {_ie} xp({z_i})) \approx {\max _i}{z_i}$. For example, $\log P_{A \oplus {C_I}}$ can be expressed as (\ref{equ:logpnc_p1_sim}).

Note that after simplification, the BS does not need to estimate the noise variance $\sigma^2$ in (\ref{equ:logpnc_p1_sim}). The physical meaning of (\ref{equ:logpnc_p1_sim}) can be understood to be selecting one constellation point with the minimum Euclidean distance among all symbols $({x_A},{x_B},{x_C})$ in set ${\chi _{{x_{A \oplus {C_I}}} = 1}}$ for computing $\log P_{A \oplus {C_I}}$ (similarly, select one constellation point in ${\chi _{{x_{A \oplus {C_I}}} =-1}}$ for computing $\log Q_{A \oplus {C_I}}$). We refer to this method as reduced-constellation demodulation scheme (details of this method can be found in \referred{MIMONCMA_Globecom}\cite{MIMONCMA_Globecom}). After that, we substitute $\log P_{A \oplus {C_I}}$ and $\log Q_{A \oplus {C_I}}$ into (\ref{equ:llr}) to obtain the LLR. The demodulation from $x_A[k] \oplus x_C^I[k]$ to ${v_A}[k] \oplus v_{{C_I}}[k]$ is a one-to-one mapping (see (\ref{equ:bpsk_demo})), and the following LLR relationship always holds
\begin{align}
LLR({v_A}[k] \oplus v_{{C_I}}[k]){\rm{ = }}LLR(x_A[k] \oplus x_C^I[k]).
\label{equ:pnc_demo}
\end{align}

We remark that the reduced-constellation demodulation scheme can be applied to MUD demodulators as well, e.g., to calculate  ${\{ {v_A}[n]\} _{n = 1,2, \dots}}$ or ${\{ {v_B}[n]\} _{n = 1,2, \dots}}$. Specifically, we refer to such an MUD decoder in NCMA as reduced-constellation MUD decoder (RMUD), to differentiate the SIC decoder used in conventional SIC-based NOMA systems.

\section{Experimental Results}\label{sec:experiments}
To evaluate the performance of our NCMA systems on weak users, especially the symbol-splitting rate-diverse NCMA system, we implemented it on software-defined radios. Section \ref{sec:experiments1} presents the experimental setup and implementation details, and Section \ref{sec:experiments2} presents the overall system performance. We remark that part of the experimental evaluations have been presented in previous sections together with the theoretical analyses to show individual users' performance of different schemes (e.g., Fig. \ref{fig:throughput_sic_noma} shows SIC does not work well in power-balanced NOMA, Fig. \ref{fig:throughput_rate_homo} presents the straightforward rate-homogeneous NCMA, and Fig. \ref{fig:throughput_rate_diverse} further evaluates the advanced rate-diverse NCMA). In Section \ref{sec:experiments2}, we focus on the system level performance of multiuser NCMA. All the experiments presented in this paper adopt the same setup as described in Section \ref{sec:experiments1}.

\subsection{Experimental Setup and Implementation Details}\label{sec:experiments1}
The rate-diverse NCMA system was built on the USRP hardware and the GNU Radio software with the UHD hardware driver. We extended the rate-homogeneous NCMA system in \referred{NCMA1,MIMONCMA_Globecom}\cite{NCMA1,MIMONCMA_Globecom} as follows: 
\begin{itemize}\leftmargin=0in
\item [a)] We modified the transceiver design in \referred{MIMONCMA_Globecom}\cite{MIMONCMA_Globecom} to support three users in addition to two users;
\item [b)] We modified the conventional rate-1/2 $[133,171]_8$ convolutional encoding and modulation scheme to the symbol-splitting encoding for QPSK packets so as to enable PNC decoding among different modulations, as discussed in Section \ref{sec:ratediverseNCMA2};
\item [c)] We implemented the XOR-CD and RMUD decoders in symbol-splitting rate-diverse NCMA as described in Section \ref{sec:ratediverseNCMA3}.
\end{itemize}

\begin{figure}
\centering
\includegraphics[width=0.6\textwidth]{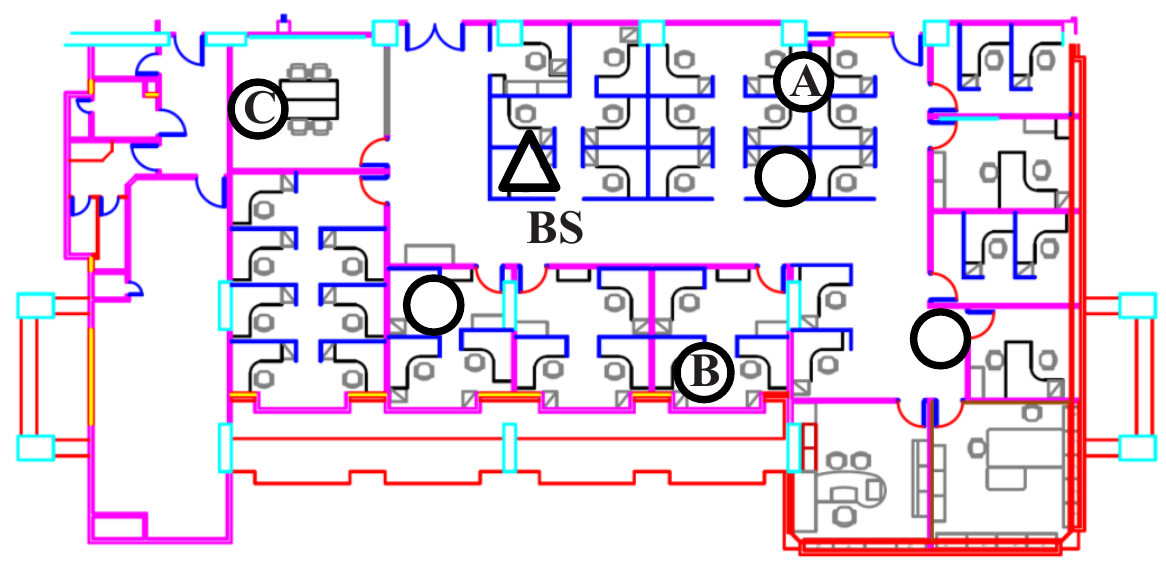}
\caption{Experimental testbed layout: the triangle represents the base station (BS) and circles represent end users. We focus on grouping three users A, B and C for NOMA/NCMA transmission after identifying and grouping active users.}
\label{fig:testbed_layout}
\vspace{-0.1in}
\end{figure}

For experimentation, we adopted the USRP hardware and the GNU Radio software with the UHD hardware driver. We deployed USRP N210s with SBX daughterboards in an indoor environment to emulate a small cell 5G network, and the topology is shown in Fig. \ref{fig:testbed_layout}. Each end user was one USRP connected to a PC through an Ethernet cable, and the BS had two USRPs connected through one MIMO cable so that the BS behaves like one node with two antennas.

Our experiments focused on NCMA transmissions. For the uplink channel, the BS sent beacon frames to trigger the end users' simultaneous transmissions\footnote{Note that we use beacons here for experimental convenience only. In an actual time slotted system, after the BS can identify and cluster the active users in Fig. \ref{fig:testbed_layout} into several groups (see Appendix \ref{sec:NCMA_RAG} for details), it will inform the groups as to the time slots when they can transmit, and the end users of a group will then transmit in their allocated time rather than being triggered by beacons of the BS.}. We examined the NCMA transmissions of one group with three end users, e.g., we clustered the three users A, B and C in Fig. \ref{fig:testbed_layout} into a group. Our experiments were carried out at 2.585GHz center frequency with 5MHz bandwidth. We performed controlled experiments for different received SNRs, and calculated SNRs using the method in \referred{HalperinSNR10}\cite{HalperinSNR10}. The received powers of signals from users A and B at the BS were adjusted to be approximately balanced at 8dB (we remark that the powers of each user could be slightly different due to channel fading, and the SNR presented here is the average SNR of all the received packets). For user C, we varied the SNR values from 8 to 14dB. For each SNR, the BS sent 1,000 beacon frames (e.g., time slots) to trigger simultaneous transmissions of the three users.

To evaluate weak users' performance and benchmark our symbol-splitting rate-diverse NCMA system, we considered the following four systems:
\begin{enumerate}\leftmargin=0in
\item \emph{SIC-based NOMA system} \\
This is the benchmarked SIC-based NOMA system with SIC decoders only. All the three users BPSK modulation. The strongest user is decoded first (i.e., user C in our experiments) and the decoding of subsequent users depends on whether the strongest user is decoded successfully or not. In Section \ref{sec:NCMAoverview5}, we focus on the throughput of the strongest user only and have seen that user C has low throughputs due to large inter-user interference.

\item \emph{Rate-homogeneous NCMA system} \\
This is the system based on the previous MIMO-NCMA system \referred{MIMONCMA_Globecom}\cite{MIMONCMA_Globecom} and it serves as a benchmark here. We extend the system in \referred{MIMONCMA_Globecom}\cite{MIMONCMA_Globecom} to support three users. The three-user PNC and MUD decoders discussed in Section \ref{sec:NCMAoverview3} are used here. PHY-layer and MAC-layer bridgings are performed to increase system throughputs. Both BPSK and QPSK modulations are considered. The three users either all use BPSK or all use QPSK.

\item \emph{Direct Rate-diverse NCMA system (DR-NCMA)} \\
This is the rate-diverse NCMA system directly generalized from the rate-homogeneous system (DR-NCMA). We study the case where two users A and B adopt BPSK, and one user C adopts QPSK. As discussed in Section \ref{sec:ratediverseNCMA1}, DR-NCMA has only one PNC decoder (to decode $C^A \oplus C^B$). In particular, only users A and B are involved in PHY-layer and MAC-layer bridgings, and the performance of user C depends on the MUD decoder.

\item \emph{Symbol-splitting Rate-diverse NCMA system (SR-NCMA)} \\
This is the rate-diverse NCMA system with symbol-splitting encoding for high-order modulated packets (SR-NCMA). Users A and B use BPSK, and user C uses QPSK. We implemented the symbol-splitting encoding scheme and the corresponding SR-NCMA PHY-layer decoders. All users can exploit PHY-layer and MAC-layer bridgings to improve throughputs.
\end{enumerate}

\subsection{Experiment Results}\label{sec:experiments2}
We evaluate the total system throughput. In Section \ref{sec:NCMAoverview5}, we already showed that rate-homogeneous NCMA outperforms SIC-based NOMA substantially. In this section, we focus on the throughput evaluations of rate-homogeneous NCMA and rate-diverse NCMA. We present the detailed PHY-layer and MAC-layer performances of the two rate-diverse NCMA systems, namely, SR-NCMA and DR-NCMA.


For the calculation of overall system throughputs, we normalize one QPSK packet to two BPSK packets. The normalized throughputs for the whole NCMA system $T{h^{sys}}$ is defined as the sum of all users' throughputs:
\begin{align}
&T{h^s} = \frac{{{L_s} \times {N_s}}}{{{N_{slot}}}},{\rm{   }}s \in \{ A,B,C\}, \\
&T{h^{sys}} = \sum\limits_{s \in \{ A,B,C\} } {T{h^s}} ,
\label{equ:throughput}
\end{align}
where ${N_s}$ is the number of messages of user $s$ that have been recovered. ${N_{slot}}$ is the number of time slots, and ${L_s}$ is the number\footnote{In NCMA, the MAC-layer RS code's parameter $L$ (see Section \ref{sec:NCMAoverview2}) can be different for different users. We choose an asymmetric case where ${L_C} = 2{L_B} = 4{L_A} = 32$ to achieve a better MAC-layer bridging performance. The asymmetric choice is preferable that was established by our prior experimental results. Detailed explanation and justification can be found in  \referred{NCMA1}\cite{NCMA1}.} of normalized BPSK packets the BS needs in order to decode message ${M^s}$.

\subsubsection{Throughput comparisons between Rate-diverse NCMA and Rate-homogeneous NCMA}\label{sec:experiments21}
\begin{figure}
\centering
\includegraphics[width=0.5\textwidth]{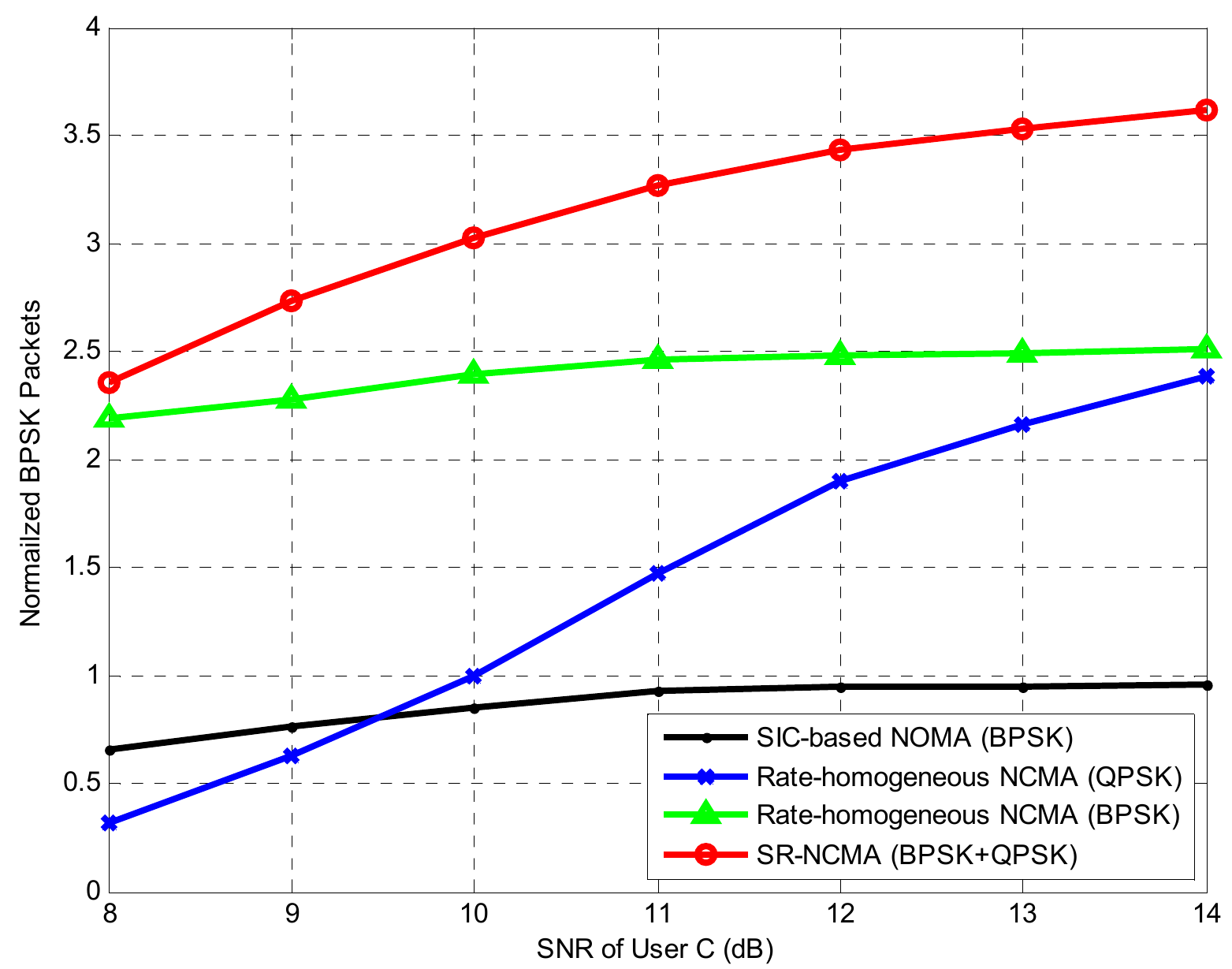}
\caption{Total system's normalized throughputs of SIC-based NOMA, rate-homogeneous NCMA and rate-diverse NCMA (SR-NCMA). The SNRs of A and B are fixed at 8dB while the SNR of C varies from 8dB to 14dB. }
\label{fig:throughput_sum}
\vspace{-0.15in}
\end{figure}

We now compare the total system throughputs of rate-diverse NCMA and rate-homogeneous NCMA, as shown in Fig. \ref{fig:throughput_sum}. As will be presented in Section \ref{sec:experiments22}, SR-NCMA performs better than DR-NCMA. Here, we use SR-NCMA as the representative of rate-diverse NCMA for comparison.

Let us focus on rate-homogeneous NCMA first. Recall that for individual users, we have seen from Fig. \ref{fig:throughput_rate_homo}(a) that when QPSK is used, users A and B have low throughputs because of their low SNRs (fixed at 8dB while the SNR of user C varies from 8dB to 14dB). But when BPSK is used, while the throughputs of A and B improve, the throughput of user C is upper-bounded by one BPSK packet per time slot as SNR increases (i.e., user C is forced to use a low modulation order, not leveraging its high SNR to obtain better throughput), as shown in Fig. \ref{fig:throughput_rate_homo}(b). Overall, although rate-homogeneous NCMA can achieve substantially throughput improvement over SIC-based NOMA, the total system throughput of rate-homogeneous NCMA systems are bounded to no more than 3 normalized BPSK packets, as shown in Fig. \ref{fig:throughput_sum}.

In contrast, rate-diverse NCMA, namely SR-NCMA, can allow users to choose their modulation order based on their channel condition. We have seen in Fig.  \ref{fig:throughput_rate_diverse}(b) that the throughputs of BPSK users A and B are comparable to those in BPSK rate-homogeneous NCMA; at the same time, user C can achieve one QPSK packet per time slot (equivalent to two BPSK packets per time slot) at high SNR. Overall, SR-NCMA achieves the highest total system throughput as shown in Fig. \ref{fig:throughput_sum}. For example, when user C's SNR is 12dB, the throughput of SR-NCMA is higher than those rate-homogeneous NCMA systems operated with BPSK and QPSK by 40\% and 80\%, respectively. In other words, rate-diverse NCMA can fully exploit the varying SNRs among weak users to improve their throughputs.

\subsubsection{Throughputs of SR-NCMA and DR-NCMA}\label{sec:experiments22}
\begin{figure}
\centering
\includegraphics[width=0.95\textwidth]{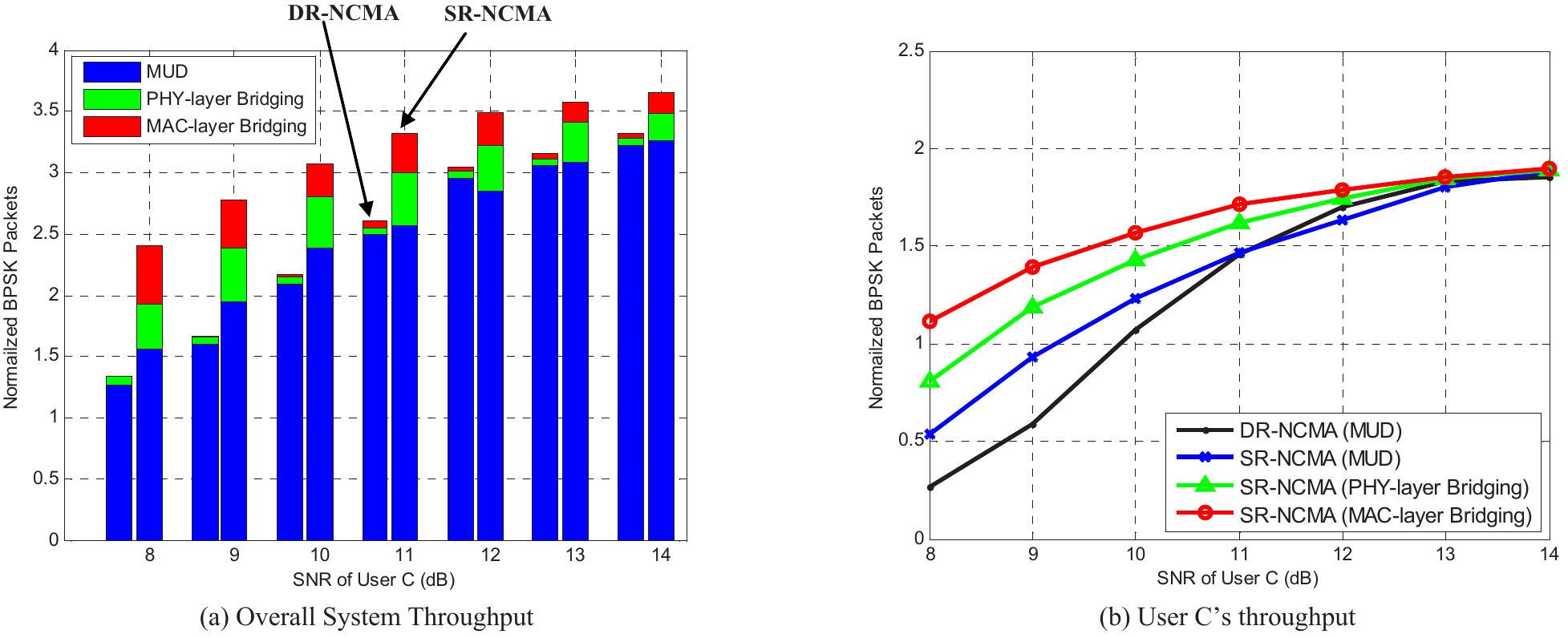}
\caption{Rate-diverse NCMA throughput comparisons: SR-NCMA versus DR-NCMA: (a) User C's throughput; (b) Total system throughput: the blue bars represent the throughputs when only MUD decoders are used; the green and red bars represent the extra throughputs when PNC decoders are also used with PHY-layer bridging and MAC-layer bridging, respectively. The SNRs of A and B are fixed at 8dB while the SNR of C varies from 8dB to 14dB. }
\vspace{-0.1in}
\label{fig:throughput_rate_diverse_detail}
\end{figure}

We now compare the system throughputs of two rate-diverse NCMA schemes in detail, namely, SR-NCMA and DR-NCMA. We first evaluate the throughputs when only MUD decoders are used. Then, we consider PNC decoders and the overall throughputs with PHY-layer and MAC-layer bridgings. The performance details of the overall system throughput are shown in Fig. \ref{fig:throughput_rate_diverse_detail}(a). To highlight the QPSK user C's throughput gain by using PNC in SR-NCMA, we also detail its performances in Fig. \ref{fig:throughput_rate_diverse_detail}(b).

\textbf{(i) Throughput Performance by MUD:}
\textbf{(i) Throughput Performance by MUD:} In Fig. \ref{fig:throughput_rate_diverse_detail}(a), the blue bars represent the overall system throughputs of SR-NCMA and DR-NCMA when only MUD decoders are used. For MUD decoders, a key difference between the two schemes is the decoding of the QPSK packets of user C (e.g., user C's MUD performances are shown in Fig. \ref{fig:throughput_rate_diverse_detail}(b)). In DR-NCMA, one MUD decoder tries to decode the whole QPSK packet $C^C$; while in SR-NCMA, two MUD decoders try to decode packets $C^{{C_I}}$ and $C^{{C_Q}}$. When user C's SNR is low, e.g., 8dB in Fig. \ref{fig:throughput_rate_diverse_detail}, it is likely that DR-NCMA fails to decode the whole QPSK packet $C^C$, but it is possible for SR-NCMA to decode one of the two packets $C^{{C_I}}$ or $C^{{C_Q}}$. Hence, we see that the MUD performance of SR-NCMA is better than DR-NCMA when user C has low SNRs. As user C's SNR increases, the MUD performances of these two schemes converge (e.g., see the blue and the black curves in Fig. \ref{fig:throughput_rate_diverse_detail}(b)).

\textbf{(ii)	Throughput Gain by PNC: PHY-layer Bridging and MAC-layer Bridging:}
A distinguishing feature of NCMA is the use of PNC packets to improve system throughput by PHY-layer bridging and MAC-layer bridging. We now evaluate the extra throughput gain due to PHY-layer bridging. Since DR-NCMA has only one PNC decoder that decodes $C^A \oplus C^B$ and the QPSK user is not involved in PNC decoding, we can see from the green bars in Fig. \ref{fig:throughput_rate_diverse_detail}(a) that PHY-layer bridging yield little improvement in DR-NCMA (i.e., PHY-layer bridging can only happen between users A and B). However, for SR-NCMA, thanks to \emph{symbol-splitting encoding}, PHY-layer bridging can also happen between the QPSK user C and the BPSK users A and B, thus improving all the three users' throughputs, e.g., the green curve in Fig. \ref{fig:throughput_rate_diverse_detail}(b) shows the performance improvement of user C. With PHY-layer bridging, SR-NCMA can have around 17\% overall system throughput improvement over that with MUD decoders only, as shown in Fig. \ref{fig:throughput_rate_diverse_detail}(a).

We finally evaluate the throughput gain due to MAC-layer bridging (see the red bars in Fig. \ref{fig:throughput_rate_diverse_detail}(a) and the red curve in Fig. \ref{fig:throughput_rate_diverse_detail}(b)). Similar to the performance gain by PHY-layer bridging, MAC-layer bridging improves the system performance of DR-NCMA very little because of the lack of PNC packets. By contrast, MAC-layer bridging can further improve the overall system throughput of SR-NCMA by around 12\% (over the throughput with PHY-layer bridging only). Therefore, the total system throughput of SR-NCMA is 40\% over that of DR-NCMA on average, as shown in Fig. \ref{fig:throughput_rate_diverse_detail}(a). The high throughput improvements using SR-NCMA indicate that SR-NCMA is a preferable solution to boost the throughputs of weak users in NOMA systems.

\section{Conclusion}\label{sec:conclusion}
We have developed a three-user NCMA system to demonstrate a practical solution for power-balanced and near power-balanced NOMA. Specifically, we put forth a \emph{rate-diverse} NCMA system wherein different users can use different signal modulations commensurate with their respective channel SNRs.

Because of the large inter-user interference, conventional SIC-based NOMA leads to low throughput under the power-balanced or near power-balanced scenarios.  We showed that, thanks to the joint use of PNC and MUD, \emph{rate-homogeneous} NCMA can already achieve substantial throughput improvements over SIC-based NOMA.

We further put forth a rate-diverse NCMA scheme to better exploit the varying SNRs among weak users under near power-balanced scenarios. A challenge for rate-diverse NCMA is the design of channel-coded PNC. This paper is the first attempt to design channel-coded rate-diverse PNC to ensure the reliability of the overall NCMA system. A key technique conceived by us to enable channel-coded rate-diverse PNC is \emph{symbol-splitting encoding}. Experimental results on our software-defined radio prototype indicate that rate-diverse NCMA can achieve higher overall system throughput in real wireless environment than rate-homogeneous NCMA. Specifically, the system throughput of rate-diverse NCMA with BPSK+QPSK modulations outperforms those of rate-homogeneous NCMA where all weak users adopt BPSK and all users adopt QPSK by 40\% and 80\%, respectively. Overall, rate-diverse NCMA is a practical solution to boost the throughput of near power-balanced NOMA systems.

\appendices
\section{NCMA Random Access and Grouping procedure (NCMA-RAG)} \label{sec:NCMA_RAG}
In our NCMA Random Access and Grouping procedure (NCMA-RAG), the base station (BS) identifies active users and clusters them into several groups to reduce the decoding complexity at PHY-layer and MAC-layer decodings. This appendix describes a contention-based random access procedure that makes use of the Zadoff-Chu (ZC) sequences for user identification. After that, NCMA user grouping is performed based on users' SNRs.

When multiple users want to transmit messages to the BS in a random access manner, the BS should have the ability to identify active users' random access requests. Here, as an example, we consider the use of the ZC sequences in LTE to identify active users. We briefly introduce the main idea of a contention-based NCMA user identification process, and more details about LTE random access procedure can be found in \referred{4GLTEBroadband}\cite{4GLTEBroadband}. NCMA-RAG consists of four main steps as follows (see Fig. \ref{fig:ncma_rag}):

\begin{figure}
\centering
\includegraphics[width=0.4\textwidth]{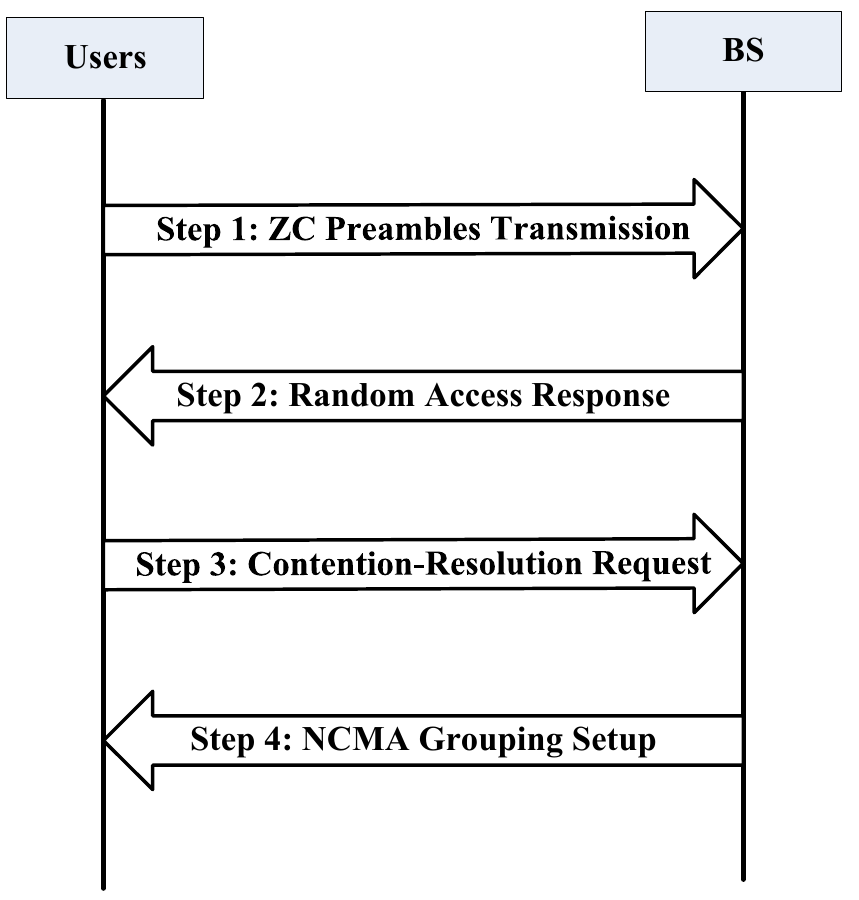}
\caption{Illustration of NCMA Random Access and Grouping procedure (NCMA-RAG).}
\label{fig:ncma_rag}
\end{figure}

\noindent \textbf{Step 1} \emph{Preamble transmission from active users to BS:} To initiate a random access, each active user randomly generates one of the available preambles and sends it to the BS. The preamble sequences are generated from cyclic shifts of the root Zadoff-Chu (ZC) sequence \referred{4GLTEBroadband}\cite{4GLTEBroadband}. The formula that generates the ZC sequence is
\begin{align}
{x_u}(m) = {e^{ - j\frac{{\pi um(m + 1)}}{{{N_{ZC}}}}}},0 \le m \le {N_{ZC}} - 1
\label{equ:zc_formula}
\end{align}

where ${N_{ZC}}$ is the length and u is the root of the ZC sequences. Based on the root ZC sequence ${x_u}(0)$, $\left\lfloor {{N_{ZC}}/{N_{CS}}} \right\rfloor$ cyclically shifted sequences are obtained by cyclic shifts of  ${N_{CS}}$ each. Each active user will choose one of the $\left\lfloor {{N_{ZC}}/{N_{CS}}} \right\rfloor$ sequences and transmit. After sending preambles, active users wait for a random access response from the BS.

\begin{figure}
\centering
\includegraphics[width=0.5\textwidth]{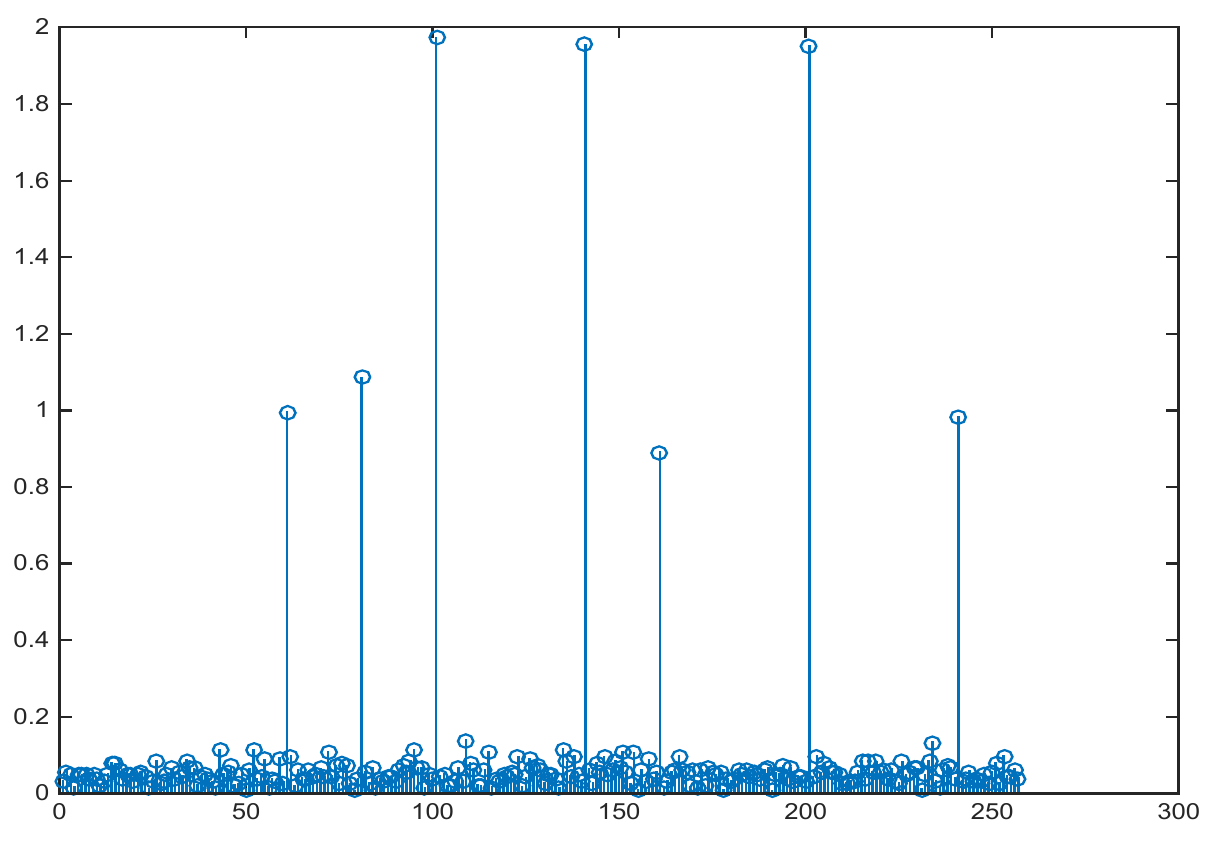}
\caption{Circular-correlation of signals consisting of superpositions of 10 ZC sequences and the root ZC sequence. In this example, the ZC sequence has length $N_{ZC}$ = 257 and root $u$ = 1. A total of 12 cyclically shifted sequences are obtained by cyclic shifts of $N_{CS}$=20 each. Two users select the ZC sequence with shift 100, two users select the ZC sequence with shift 140, and two users select the ZC sequence with shift 200.  }
\label{fig:zc_correlation}
\end{figure}

\noindent \textbf{Step 2} \emph{Random access response from BS to active users:} After receiving preambles from active users, the detection of random access preambles is based on the correlation of the received signal and the root ZC sequence. Fig. \ref{fig:zc_correlation}) shows an example of a circular correlation of signals consisting of superpositions of 10 ZC sequences and the root ZC sequence. The BS can estimate the number of active users from the number of peaks, and the indices of the ZC preambles from the positions of the peaks (the shifted versions of the root ZC sequence).

The random access response from the BS to active users contains the index of the random access preamble sequence. Based on different indices, the random access responses are sent in a TDMA manner. As long as the active users use different preambles (i.e., different indices), from the downlink response signaling it is clear from the index that which user has been identified by the BS. However, the identification may not be unique since there is a certain probability of contention. That is, multiple users may choose the same preamble in Step 1. In this case, multiple users will receive the same downlink response and the same preamble index (e.g., in Fig. \ref{fig:zc_correlation}), two users select the ZC sequence with shift 100, two users select the ZC sequence with shift 140, and two users select the ZC sequence with shift 200). Resolving these contentions is part of the subsequent steps.

\noindent \textbf{Step 3} \emph{Contention-resolution request from active users to BS:} After receiving the random access response from the BS, users decode the preamble index embedded in the response to see if the index is the same as what they sent in Step 1. For each preamble index (random access response), the user(s) who decoded the same preamble index as that in Step 1 will send a contention-resolution request with its user identity (e.g., a user-identified number, or a random number) to the BS (e.g., users with different indices will transmit requests to the BS in different time slots in Step 3). This identity is used for contention resolution. That is, there is no collision if only one user adopted this preamble index in Step 1; however, if more than one user adopted the same preamble index (e.g., the two users select the ZC sequence with shift 100 in Step 1 in Fig. \ref{fig:zc_correlation}), they will transmit the contention-resolution requests at the same time but with different user identities, leading to a collision.

\noindent \textbf{Step 4} \emph{NCMA grouping setup from BS to active users:} This step is also referred to as the contention resolution stage. If there is no collision in Step 3, the BS will receive the contention-resolution request and estimates the user's SNR. After collecting all (non-collided) active users' SNRs, the BS clusters users into several groups based on the two key design decisions put forth in this paper (i.e., see Section \ref{sec:intro}, the strong user operation and the weak user operation). The BS sends back the grouping decisions to the users. The users wait for NCMA transmission.

However, the collided users, e.g., those sent the contention-resolution requests at the same time in Step 3, will not receive acknowledgments due to collisions. In this case, they need to restart the procedure from the first step (e.g., the collided users will randomly choose their ZC preambles again and transmit in Step 1 \referred{4GLTEBroadband}\cite{4GLTEBroadband}).

\ifCLASSOPTIONcompsoc
  \section*{Acknowledgments}
\else
  \section*{Acknowledgment}
\fi

The work of H. Pan and S. C. Liew was supported by the General Research Funds (Project No. 14204714) established under the University Grant Committee of the Hong Kong Special Administrative Region, China.
The work of L. Lu was partially supported by the NSFC (Project No. 61501390) and partially supported by AoE grant E-02/08, established under the University Grant Committee of the Hong Kong Special Administrative Region, China.

\bibliographystyle{IEEEtran}
\bibliography{jsac-ncma}

\end{document}